\documentclass{article}
\usepackage{amsmath,amsfonts,amssymb}
\usepackage{authblk}
\usepackage{graphicx}
\usepackage{braket}
\usepackage{array}
\usepackage[a4paper, margin=1in]{geometry}
\usepackage{hyperref}

\usepackage{xcolor}
\usepackage{soul}
\usepackage{csquotes}

\title{Time Interfaces Enable Free-Space Excitation of Frozen Electromagnetic Waves and Topological Edge States in Graphene}

\author{
	Shivaksh~Rawat$^{1,\ast}$, Samyobrata Mukherjee$^{1}$, Jeffery Allen$^{2}$, Monica Allen$^{2}$,	Gennady~Shvets$^{1,\dagger}$\and
	\small$^{1}$School of Applied and Engineering Physics, Cornell University, Ithaca, NY 14853, USA.\and
    \small$^{2}$Air Force Research Laboratory, Air Warfare Directorate, Eglin AFB, Florida 32542, USA.\and
	\small Email: sr939@cornell.edu$^\ast$, gshvets@cornell.edu$^\dagger$
}

\date{}
\begin{document}
\maketitle
\begin{abstract}
Time-varying two-dimensional materials, such as graphene under ultrafast photoexcitation, offer a promising platform for engineering temporal interfaces for electromagnetic waves. Rapidly promoting valence electrons to the conduction band generates a high-energy, non-thermal hot-electron population that subsequently thermalizes with the pre-existing thermal electron population. We study the excitation of plasmonic modes in graphene by the time interface created by ultrafast photoexcitation using a graphene electron-transport model self-consistently coupled to electromagnetic simulations. We find that, in addition to dynamic, propagating plasmonic modes, the time interfaces excite static, spatially varying electromagnetic modes in the interaction region, sustained by streaming transverse electrons. We also demonstrate a novel technique for coupling free-space light directly to highly localized topological edge states at the magnetic-inversion boundary on graphene, eliminating the need for near-field excitation techniques.
\end{abstract}

\section{Introduction}
\label{Sec:1}

Time-varying optical materials have attracted significant interest over the last decade because they offer a novel paradigm for photonics \cite{Pendry2022PTVM, Yu2026TVMReview, Review2026AdvSci}. Time-varying photonic materials provide a platform to explore novel physical phenomena such as photon acceleration \cite{Shcherbakov2019PA}, expanding momentum bandgaps in photonic time crystals \cite{Asadchy2024NatPhoto, Salehi2026nonlocalPTC} including proposals for their realization using bound states in the continuum \cite{P_Garg_SciAdv2026, Puneet2026BIC_MM}, negative wave extinction \cite{Shcherbakov2019NE}, and nanoscale magnetization arising from rapid free-carrier generation in semiconductors \cite{S_Rawat_AdvSci2026}. Experiments with time-varying photonic media have demonstrated time diffraction at optical frequencies in transparent conducting oxide films \cite{Tirole2023TD}, time reflection in transmission-line metamaterials \cite{Moussa2023TLM}, and plasmonic time crystals in metamaterials \cite{PlasmonicMTC_Nat_2026}.

Graphene, a two-dimensional material comprising an atomic monolayer of carbon atoms arranged in a hexagonal lattice, has been intensely investigated since its isolation two decades ago because of its unique electrical and optical properties \cite{Geim2009RevModPhys, Koppens2011Nano}. Graphene supports highly confined, deeply sub-wavelength, tunable, low-loss plasmonic states \cite{Woessner2015, Ni2018}, making it a suitable platform for tailoring strong light-matter interactions. While graphene's optical properties are typically tuned using electrostatic gating, which is a relatively slow process compared to the typically far-IR graphene plasmons (GPs), another pathway to modulate the optical response of graphene involves the ultrafast photoexcitation of graphene electrons from the valence to the conduction band \cite{Malic2011PRB, Tomadin2013PRB, Mihnev2015NatComms, Klaas2021RSE}. This method enables the creation of a sharp photonic time interface (TI) for GPs \cite{Bakunov2019PRB, Maes2017OptLett}. This work investigates the heretofore unexplored effects of non-equilibrium electron dynamics, such as the creation of a non-thermal electron population and its subsequent thermalization, in the immediate aftermath of photoexcitation on GP excitation and propagation. We find that the TI engineered by ultrafast photoexcitation couples a free-space (FS) electromagnetic wave to various plasmonic modes in graphene.

We consider two cases for a free-standing graphene sheet: first, without an external static magnetic field bias applied perpendicular to the graphene, and second, with one. Remarkably, it is revealed that in the case of graphene without a magnetic bias (unmagnetized graphene), an s-polarized FS wave at a TI interacts with graphene electrons, exciting a zero-frequency magnetic-field mode supported by freely streaming electrons perpendicular to the plane of incidence, which cannot be excited by an oscillating source \cite{Dafei_2016_NatComms, Mukherjee2025}. In contrast, a p-polarized FS wave excites only forward- and backward-propagating GPs. In the second case of graphene biased by a spatially uniform magnetic field (magnetized bulk graphene), any FS-wave polarization at a TI excites all three modes: a static electromagnetic mode with nonzero electric and magnetic components localized in the interaction region, and forward- and backward-propagating bulk graphene magnetoplasmons (GMPs). While the generation of a static electric (magnetic) field mode in a purely capacitive (inductive) time-varying metasurface has been predicted \cite{Tretyakov2023Nanoph}, no prior work has reported exciting a static electromagnetic mode with both nonzero electric and magnetic field components at a TI. Lastly, we describe a novel technique for coupling FS light into edge GMPs at a domain edge in magnetized graphene using a TI. The external bias magnetic field flips direction at the magnetic domain edge over a finite length, creating a topological heterojunction that supports topologically protected Kelvin and Yanai-like edge GMPs in addition to non-topological low-frequency Rossby waves and Volkov-Pankratov (VP) edge GMPs \cite{Dafei_2016_NatComms, Mukherjee2025, Delplace2017topological, Volkov1985, Tchoumakov2017VP}. During photoexcitation by an intense pump pulse, s-/p-polarized FS light couples not only to a topological forward-propagating Yanai-/Kelvin-like edge GMP but also to non-topological, forward- and backward-propagating Volkov-Pankratov edge GMPs and near-zero-frequency Rossby waves. Additionally, we observe no backward-propagating topological edge GMP under this coupling, indicating complete immunity to backscattering. While topology and temporal topological edge states have been studied in the context of momentum bands in photonic time crystals \cite{Lustig2018}, to the best of our knowledge, this is the first use of TIs to excite topological edge states in frequency bandgaps, interfacing temporal photonics with topological photonics.
\begin{figure}
	\centering
	\includegraphics[width=0.8\textwidth]{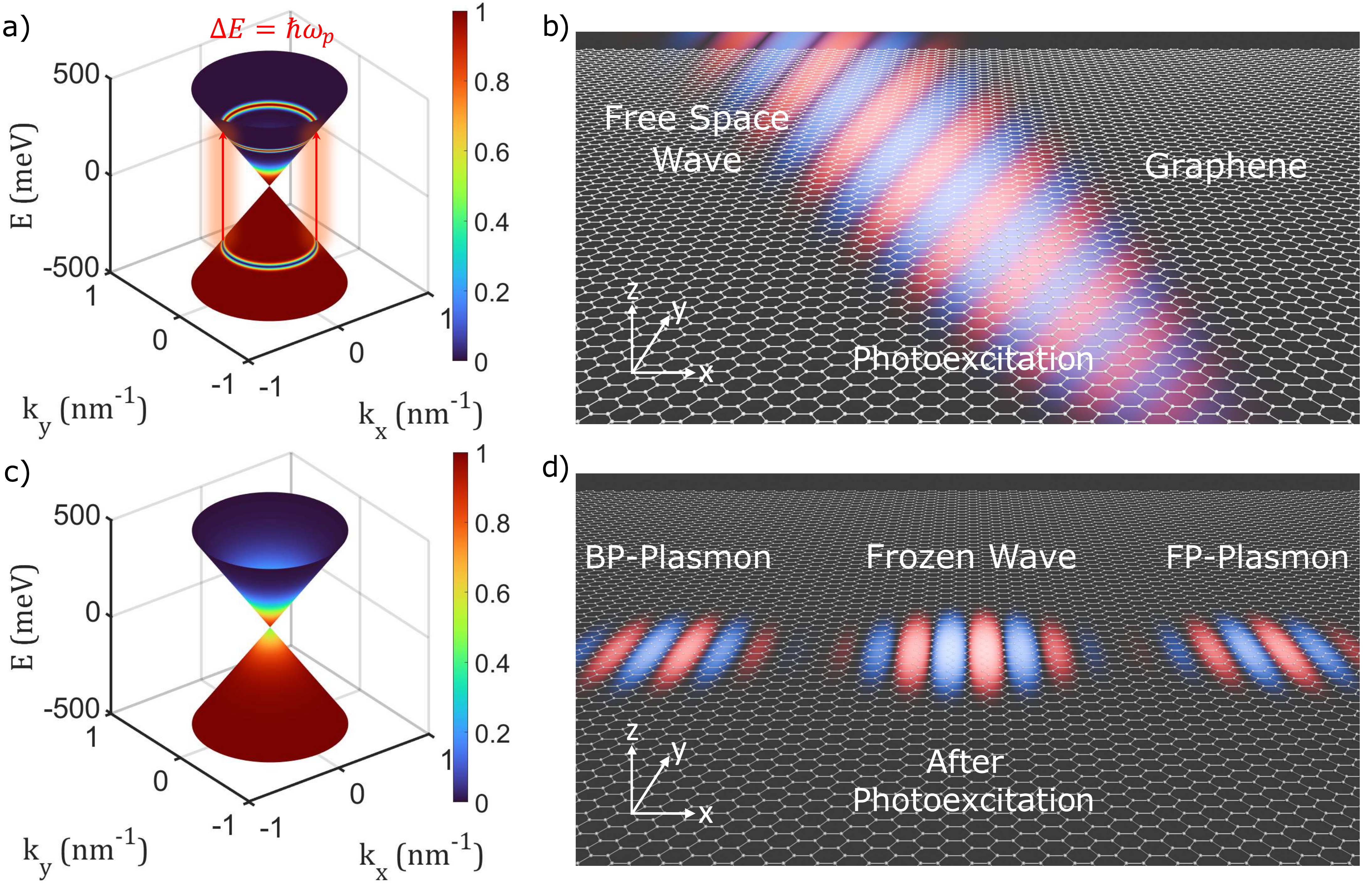}
	\caption{\textbf{Free-Space Excitation of a Frozen Wave in Graphene.}
		\textbf{a}. Graphene electronic band near a Dirac point during photoexcitation; the narrow electron band centered at $\hbar\omega_p/2$ represents non-thermal electrons. Color shows occupation. \textbf{b}. Free-space light interacting with graphene electrons at a TI created during photoexcitation. Plane of incidence: $x-z$ plane at $y=0$. \textbf{c}. Same as \textbf{a} but after thermalization of the non-thermal electrons. \textbf{d}. Formation of a frozen wave along with forward-propagating (FP) and backward-propagating (BP) plasmons at the TI.}
	\label{fig:MT_1} 
\end{figure}

The remainder of the manuscript is organized as follows. Section~\ref{Sec:2} introduces an electron-transport model that describes the electronic current density in time-varying graphene. Section~\ref{Sec:3} analyzes the bulk GP and GMP dispersion for unmagnetized and magnetized graphene, respectively, as well as edge GMP dispersion. Section~\ref{Sec:4} reports simulation results for FS light interacting with graphene during photoexcitation. Section~\ref{Sec:4.1} examines the FS excitation of a static magnetic mode in unmagnetized graphene. At a TI, we investigate the scattered plasmon amplitude and frequency as functions of the absorbed pump-photon energy. Section~\ref{Sec:4.2} demonstrates the excitation of a static electromagnetic mode in magnetized graphene at a TI. Section~\ref{Sec:4.3} describes a novel technique for coupling FS light into both topological and non-topological edge GMPs at a magnetic domain edge. Section~\ref{Sec:5} discusses the implications of the findings.

\section{Dynamic Current Density in Time-Varying Graphene}
\label{Sec:2}
An intense, ultrafast pump pulse rapidly promotes valence electrons in n-doped ($\mu_i>0$) graphene into the conduction band, as shown in Fig.~\ref{fig:MT_1}a, thereby modifying graphene's optical response and creating a sharp TI as an FS wave is incident on graphene, as illustrated in Fig.~\ref{fig:MT_1}b. The photoexcitation generates a non-thermal electron population that thermalizes primarily via electron-electron collisions, as shown in Fig.~\ref{fig:MT_1}c. The TI scatters FS light into GP modes, as shown in Fig.~\ref{fig:MT_1}d, while conserving the in-plane wave vector. Electrons in graphene exhibit linear dispersion $E=v_F|\mathbf{p}|$ near the Dirac ($K$ and $K'$) points in the Brillouin zone for electron energies significantly less than the tight-binding hopping parameter, i.e., $E\ll\gamma_0\sim2.7$ eV, where $\mathbf{p}$ is the electron momentum and $v_F=1.1\times10^6$ m/s is the Fermi velocity \cite{Geim2009RevModPhys, Wallace1947TheBT}. Although this linear dispersion can be described by the massless Dirac equation, an effective cyclotron mass can be assigned to graphene electrons as $m_e(E)=E/v_F^2$ \cite{Geim2009RevModPhys}. This implies that higher-energy non-thermal electrons possess a larger effective mass than thermal electrons, which occupy lower energies and thus have a lighter mass. Because heavier non-thermal electrons do not accelerate as readily as lighter thermal electrons in external fields, the two electron populations must be modeled separately.
\begin{figure}
	\centering
	\includegraphics[width=0.9\textwidth]{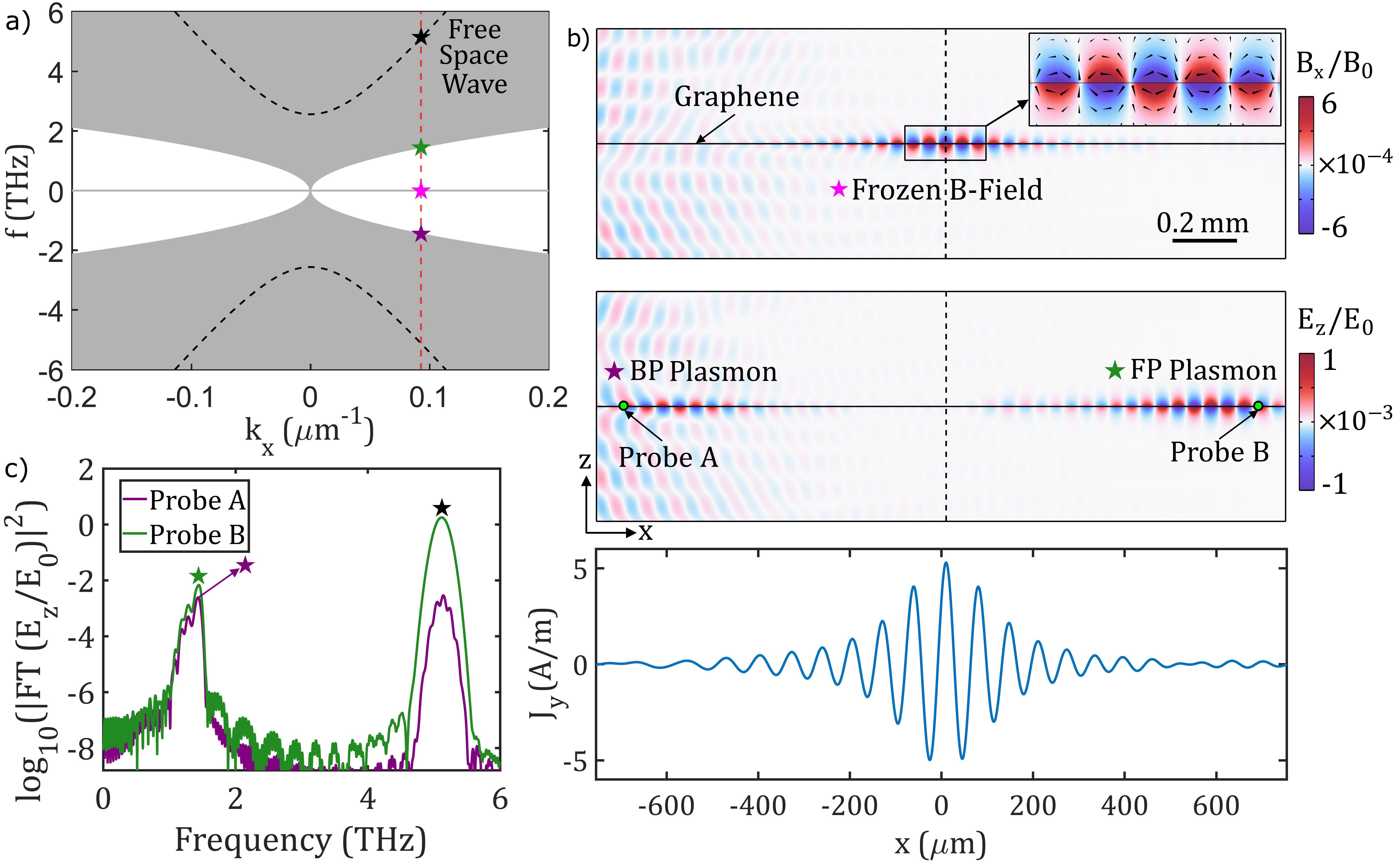}
	\caption{\textbf{Frozen B-Field in Unmagnetized Graphene.}
		\textbf{a}. Bulk-plasmon dispersion in unmagnetized graphene, with stars denoting different modes accessible at a TI; the black dashed curve shows the air light line with a $k_z\neq0$ offset, while the dashed vertical line indicates the conserved in-plane wave vector. \textbf{b}. The top panel shows the frozen $\mathrm{B}_x$ field (pink star), and the inset shows an enlarged view of the frozen wave with magnetic field lines (black cones), while the middle panel shows the $\mathrm{E}_z$ field of propagating plasmons (green and purple stars) at 17.5 ps (11 ps after the TI), excited using a mixed-polarization free-space (FS) Gaussian wave. The vertical dashed line indicates the focus of the Gaussian FS wave where it intersects graphene in the top and middle panels. The scale is the same in the top and middle panels. The bottom panel shows the y-component of the current density on graphene. \textbf{c}. Spectrum of the $\mathrm{E}_z$ field recorded at the two probe locations shown in the middle panel in $\textbf{b}$.}
	\label{fig:MT_2} 
\end{figure}

For pump photon energies satisfying $\hbar\omega_p/2<\mu+\hbar\omega_{op}$, where $\hbar\omega_{op}\approx200$ meV is the energy of an intra-valley optical phonon at the $\Gamma$ point, the non-thermal electron population relaxes into the thermal sea predominantly via electron-electron scattering \cite{TielrooijPRB2013}. Although this scattering proceeds microscopically as a cascade of elastic collisions, we approximate it as a single-step transfer of non-thermal electrons into the thermal sea, implying that the non-thermal electron energy and momentum densities share the same relaxation time constant, $\tau_\textrm{e-e}\approx50$ fs \cite{Polini2013NatComms}. Further, we assume that collisions between non-thermal electrons are rapid enough to isotropize their momenta but insufficient to fully thermalize them. Moreover, conduction-band electrons recombine with valence-band holes predominantly via Auger recombination and inter-valley optical phonon emission, with a combined time constant of several ps, and do not significantly affect electron dynamics during ultrafast thermalization \cite{Rana2007PRB}. Thus, to isolate the effect of non-thermal electron dynamics, we ignore electron-hole recombination and assume a constant conduction-band electron density after non-thermal electron generation. Therefore, while a more sophisticated model of electron dynamics in photoexcited graphene \cite{TielrooijPRB2013, Tomadin2013PRB} may yield quantitative adjustments to the rates of momentum and energy exchange, the simplified model adequately captures the physics while allowing us to probe the excitation of GPs and GMPs at a TI created by ultrafast photoexcitation.

\begin{figure}
	\centering
	\includegraphics[width=0.7\textwidth]{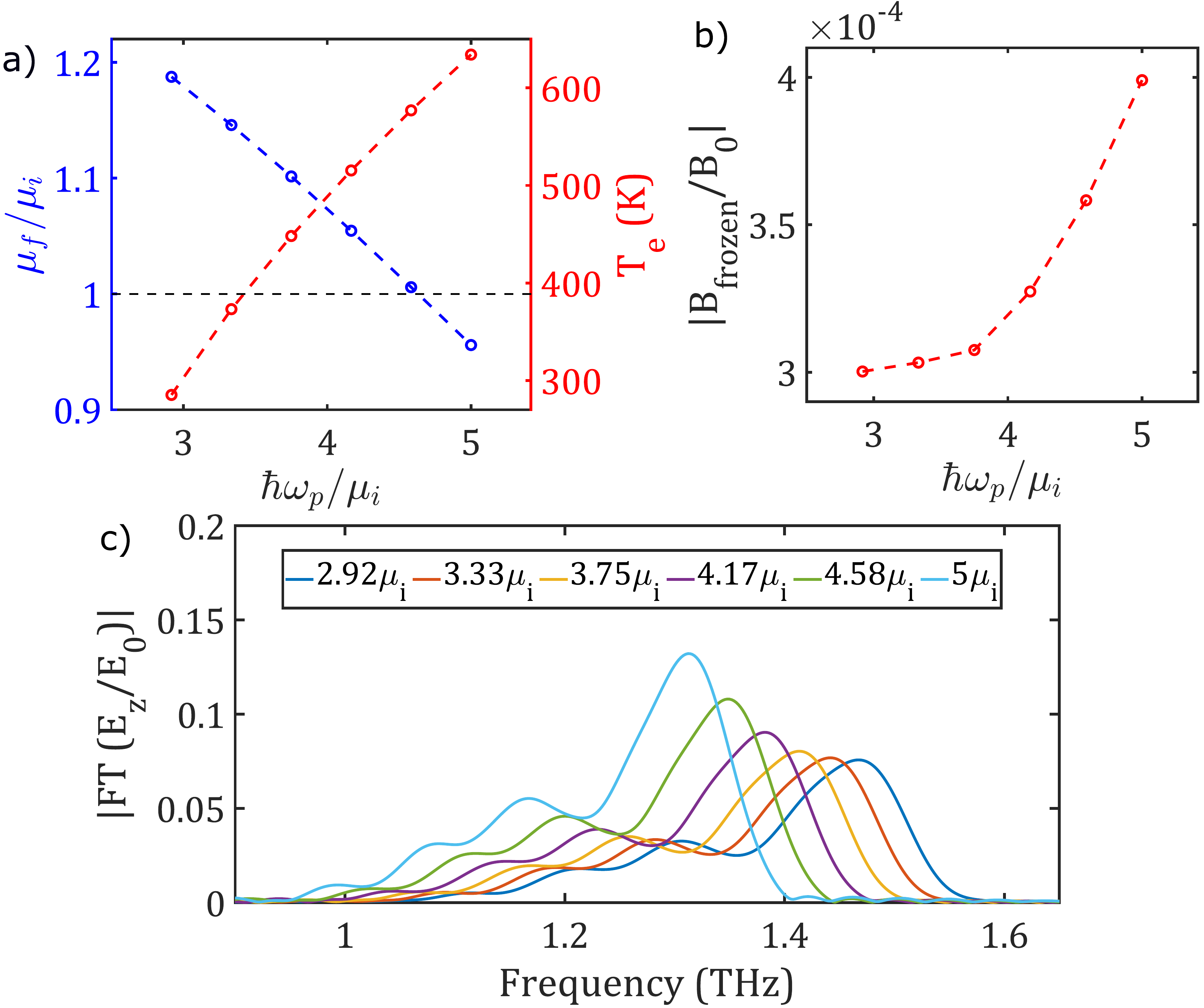}
	\caption{\textbf{Pump Photon Energy Dependence of Frozen Wave and Scattered Plasmon Amplitude.}
		\textbf{a}. Chemical potential and electron temperature for different pump photon energies. \textbf{b}. Maximum amplitude of the frozen magnetic mode as a function of photon energy. \textbf{c}. Spectra of the $\mathrm{E}_z$ signal recorded at Probe B shown in Fig.~\ref{fig:MT_2}b for different pump photon energies.}
	\label{fig:MT_3} 
\end{figure}
In electrodynamics, free-standing graphene located in the $x$-$y$ plane at $z=0$ is represented as a two-dimensional sheet that supports a surface current density, $\textbf{J}_s$, resulting in a discontinuity in the magnetic field, $\textbf{H}$, across the sheet as follows:
\begin{equation}
    \hat{z}\times\left[\textbf{H}_{z=0^+}(x,y,t)-\textbf{H}_{z=0^-}(x,y,t)\right]=\textbf{J}_s,
\end{equation}
where the total surface current density $\textbf{J}_s=\textbf{J}_T+\textbf{J}_{NT}$ is the sum of the current densities generated by the thermal and non-thermal electron populations. The time evolution of $\textbf{J}_T$ and $\textbf{J}_{NT}$ is obtained from the Boltzmann transport equation (see the Supplementary Information for a detailed derivation) and is given by
\begin{equation}
\begin{split}
    \frac{\partial \textbf{J}_{T}}{\partial t}&=\frac{e^2\mu}{\pi\hbar^2}\textbf{E}-\frac{ev_F^2}{\mu}\textbf{J}_{T}\times\textbf{B}_s+\left(\frac{E_{NT}}{\mu}\right)\frac{\textbf{J}_{NT}}{\tau_\textrm{e-e}}-\frac{\textbf{J}_T}{\tau_m}, \\
    \frac{\partial\textbf{J}_{NT}}{\partial t}&=\frac{e^2v_F^2N_{NT}}{2E_{NT}}\textbf{E}-\frac{ev_F^2}{E_{NT}}\textbf{J}_{NT}\times\textbf{B}_s-\frac{\textbf{J}_{NT}}{\tau_\textrm{e-e}}.
    \label{eq:Total_dJ_dt}
\end{split}
\end{equation}
Here, $E_{NT}$ is the average non-thermal electron energy, and we assume an effective momentum relaxation time constant, $\tau_m \approx 10$ ps \cite{Castro2010PRL}. The conduction-only, two-electron population model remains valid even if we assume a separate, thermalized quasi-Fermi distribution for valence-band holes. This is because the relatively large pre-existing conduction-electron density in strongly n-doped graphene creates a stark asymmetry between the two bands. Following ultrafast thermalization, the valence-band hole chemical potential rapidly relaxes to near the Dirac point, where the electronic density of states is very low. As a result, few states are available to support a separate photoexcited hole-current density in the valence band. In contrast, the chemical potential of the conduction-band electron population exceeds the initial chemical potential, where the density of states is substantial. Therefore, the total dynamic surface current density $\textbf{J}_s(t)$ arises entirely from conduction-band electrons, with little to no contribution from valence-band holes.
\begin{figure}
	\centering
	\includegraphics[width=0.9\textwidth]{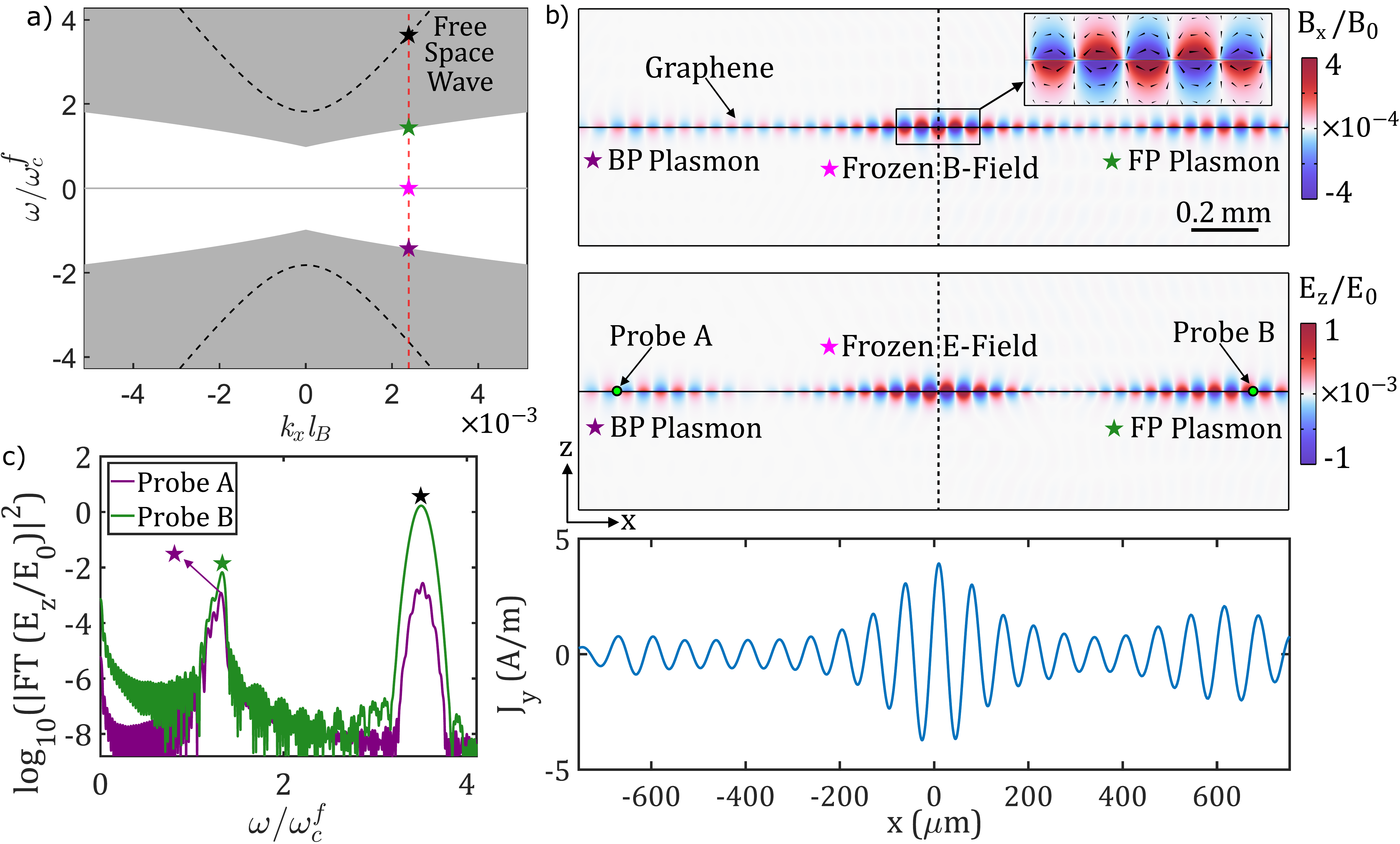}
	\caption{\textbf{Frozen Electromagnetic Field in Magnetized Graphene.}
		\textbf{a}. Bulk-plasmon dispersion of magnetized graphene; colored stars denote different modes accessible at a TI; the black dashed curve shows the air light line with a $k_z\neq0$ offset, while the dashed vertical line indicates the conserved in-plane wave vector. \textbf{b}. The top (middle) panel shows the $\mathrm{B}_x$ ($\mathrm{E}_z$) field of the frozen and propagating plasmons, and the inset shows an enlarged view of the frozen wave with magnetic field lines (black cones), while the bottom panel shows the $\textrm{J}_y$ current density at 20 ps (13.5 ps after the TI), excited using a mixed-polarization FS Gaussian pulse; the intersection of the vertical dashed line with graphene indicates the focus of the Gaussian wave. \textbf{c}. Spectrum of the $\mathrm{E}_z$ field recorded at the two probe locations shown in $\textbf{b}$, showing the conversion of FS light into propagating plasmons at a TI.}
	\label{fig:MT_4} 
\end{figure}

\section{Plasmon Dispersion in Doped Free-Standing Graphene}
\label{Sec:3}
We consider a free-standing graphene sheet in the $z=0$ plane and apply a static magnetic bias, $\textbf{B}_s=\textrm{B}_s\hat{\textbf{z}}$, perpendicular to the graphene. This leads to Landau level (LL) quantization in graphene with discrete, highly degenerate energy levels at $E_n=\mathrm{sgn}(n)\hbar v_F\sqrt{2|n|}/l_B$, where $n$ denotes the LL number and $l_B=\sqrt{\hbar(e\textrm{B}_s)^{-1}}$ is the magnetic length. The energy spacing between LLs decreases for high $n$ and low $\textrm{B}_s$, and disorder broadens the LLs. Therefore, for highly doped graphene in a weak bias magnetic field ($\textrm{B}_s \lesssim 1$ T), we may safely approximate the density of electronic states as a linear function of electron energy (as in the case of $\textrm{B}_s=0$), and the magneto-optical conductivity of graphene can be modeled accurately by the semi-classical Drude model at low frequencies \cite{LLs_Quant_2}. We consider highly doped graphene with chemical potential $\mu_i=120$ meV in a bias field $\textrm{B}_s=1$ T resulting in $n=9$ LLs being filled and a cyclotron frequency $\omega_c=e\textrm{B}_sv_F^2/\mu_i\approx 10.083\times10^{12}$ rad/s ($f_c\approx1.605$ THz). Pauli blocking prevents low-energy inter-LL transitions between valence and conduction band LLs, and the Drude model is valid for all incident light frequencies $\mathrm{f}_0\ll\left(\sqrt{10}+3\right)v_F\sqrt{\frac{e\textrm{B}_s}{2\pi^2\hbar}}\approx60$ THz.

Solving Maxwell's equations with the appropriate equations of motion for the charge carriers (Lorentz force equation and continuity equation) yields the bulk magnetoplasmon dispersion with: (i) two dispersive magnetoplasmon bands $\omega_{\pm}(k)=\pm\sqrt{\omega_c^2+\alpha^2k}$, where $k=\sqrt{(k_x^2+k_y^2)}$ is the in-plane wavevector and $\alpha=\sqrt{e^2\mu/(2\pi\hbar^2\epsilon_0)}$ and (ii) a zero frequency band with  $\omega_0(k)=0$, where $\omega_c=0$ yields the GP dispersion for unmagnetized graphene. Furthermore, we used the plane-wave expansion method to calculate the bulk and edge GMP dispersion at a domain edge in graphene \cite{Dafei_2016_NatComms, Mukherjee2025}. See the Supplementary Information for detailed band structure calculations.

\section{Results}
\label{Sec:4}
The free-space (FS) wave with electric field amplitude, $\mathrm{E}_0=10^6$ V/m, is incident on graphene in the x-z plane at $y=0$ and has a Gaussian profile in both time and space. Table~\ref{Table1} lists the parameters used in the calculations, and we assume the system is at 4 K. During the interaction between the FS wave and graphene electrons, an intense pump pulse rapidly excites electrons from the valence to the conduction band, inducing a TI. After the electrons generated during the TI thermalize (the TI lasts $T_{TI}\sim150$ fs), the total electron density reaches $1.35\times10^{12}\ \mathrm{cm}^{-2}$. The free-space wave's optical cycle is longer than the TI created by ultrafast photoexcitation ($T_0=1/f_0\approx195\ \mathrm{fs}>T_{TI}$). Therefore, the pump pulse creates a relatively sharp TI for the FS light. The following sections present results from numerical studies of TI-enabled coupling of FS light to GPs and GMPs. We find that FS light couples to i) a static magnetic field mode and bulk GPs in unmagnetized graphene, ii) a static electromagnetic mode and bulk GMPs in magnetized graphene, and iii) topological and non-topological edge GMPs at a magnetic domain edge.

\subsection{Static Magnetic Field Mode in Unmagnetized Graphene}
\label{Sec:4.1}
Fig.~\ref{fig:MT_2}a shows the bulk GP dispersion for an infinite sheet of unmagnetized graphene, where the bulk bands (gray) with $f\neq0$ represent dynamic, topologically trivial, propagating transverse magnetic (TM) GP modes, and the zero-frequency mode corresponds to a static, spatially varying GP mode comprising only magnetic fields (with zero electric-field amplitude). The dynamic or static GP can be selectively excited by controlling the polarization of the incident FS wave. A mixed-polarization Gaussian wave is considered, i.e., $\hat{\textbf{E}}_0=\left(\frac{1}{\sqrt{2}}\right)\left(\ket{s}+\ket{p}\right)$, where $\hat{\textbf{E}}_0$ is the incident polarization and $\ket{s}$/$\ket{p}$ represents s-/p-polarization, with a pulse width of 2 ps, centered at 5 ps, incident at $\theta=60^\circ$ relative to $\hat{z}$ to excite all possible GPs. The black dashed curve in Fig.~\ref{fig:MT_2}a shows the light-cone dispersion relation $\omega=\pm c\left(k_x^2+k_z^2\right)^{1/2}$, with $k_y=0$ and $k_z=k_0\cos{\theta}$ to account for the FS wave's incident angle. The TI is introduced at 6.5 ps, when the pulse's central maximum reaches the focus, set at the origin and shown as the intersection of the dashed vertical line with the graphene plane in Fig.~\ref{fig:MT_2}b. At the TI, a static magnetic-field mode is excited, with a spatially varying static $\mathrm{B}_x$ field shown at 17.5 ps in the top panel of Fig.~\ref{fig:MT_2}b. The inset shows the circulating magnetic field lines (black cones) in an enlarged view of the frozen wave. This zero-frequency mode is supported by a static current density with spatial variation along x, as shown in the bottom panel of Fig.~\ref{fig:MT_2}b, and results from electrons streaming freely along the y-direction, i.e., $J_y(x)\neq0$ and $\partial J_y(x)/\partial t=0$. Additionally, the middle panel of Fig.~\ref{fig:MT_2}b shows forward- and backward-propagating TM GPs excited by the time interface. Fig.~\ref{fig:MT_2}c shows the $\mathrm{E}_z$ field spectrum recorded at the two probe locations, indicated by green circles in the middle panel of Fig.~\ref{fig:MT_2}b. The redshifted peak at f $\approx$ 1.44 THz corresponds to the forward- and backward-propagating plasmons, shown as purple and green stars in Fig.~\ref{fig:MT_2}a, excited by the TI at a fixed in-plane wave vector. 
\begin{table}
\centering
 \begin{tabular}{ | m{13em} | m{2.5cm}| m{3.5cm} | } 
 \hline
 \bf{Physical Quantity} & \bf{Symbol} & \bf{Value} \\ 
 \hline
 Initial Chemical Potential & $\mu_i$ & $120$ meV \\ 
 \hline
 Pump Photon Energy & $E_p=\hbar\omega_p$ & $3.33 \mu_i=0.4$ eV \\ 
 \hline
 Static Magnetic Bias & $\textrm{B}_s$ & $1$ T \\ 
 \hline
 Free Space Wave Frequency & $\mathrm{f}_0$ & $5.12$ THz (21.2 meV) \\ 
 \hline
 Final Chemical Potential & $\mu_f$ & $137$ meV \\ 
 \hline
 Final Cyclotron Frequency & $\omega_c^f=e\textrm{B}_sv_F^2/\mu_f$ & $2\pi\times1.406$ THz \\ 
 \hline
\end{tabular}
\caption{Summary of parameters and values for graphene and external electromagnetic fields.}
\label{Table1}
\end{table}

We also examined how varying pump-photon energy affects the amplitude of GPs excited at a TI. We varied the pump-photon energy from $2.92\mu_i$ to $5\mu_i$ and calculated the scattered plasmon amplitude, keeping the number of electrons excited from the valence to the conduction band constant. Fig.~\ref{fig:MT_3}a shows that as pump photon energy increases, the electron temperature rises, while the chemical potential decreases to maintain the same conduction-electron density. Consequently, higher-energy pump photons produce larger amplitudes of the static magnetic mode, as shown in Fig.~\ref{fig:MT_3}b. This follows from the third term in the thermal current density equation in eq.~\ref{eq:Total_dJ_dt}, which suggests that as $E_{NT}$ increases and $\mu$ decreases, the ratio $E_{NT}/\mu$ increases for higher-energy photons, producing stronger current densities during photoexcitation. This implies that non-thermal electron energy strongly affects plasmon excitation and leads to stronger GPs as photon energies increase. Furthermore, because higher-energy non-thermal electrons reduce $\mu_f$ more strongly, the GP frequency redshifts as shown in Fig.~\ref{fig:MT_3}c.

\begin{figure}[t!]
	\centering
	\includegraphics[width=0.7\textwidth]{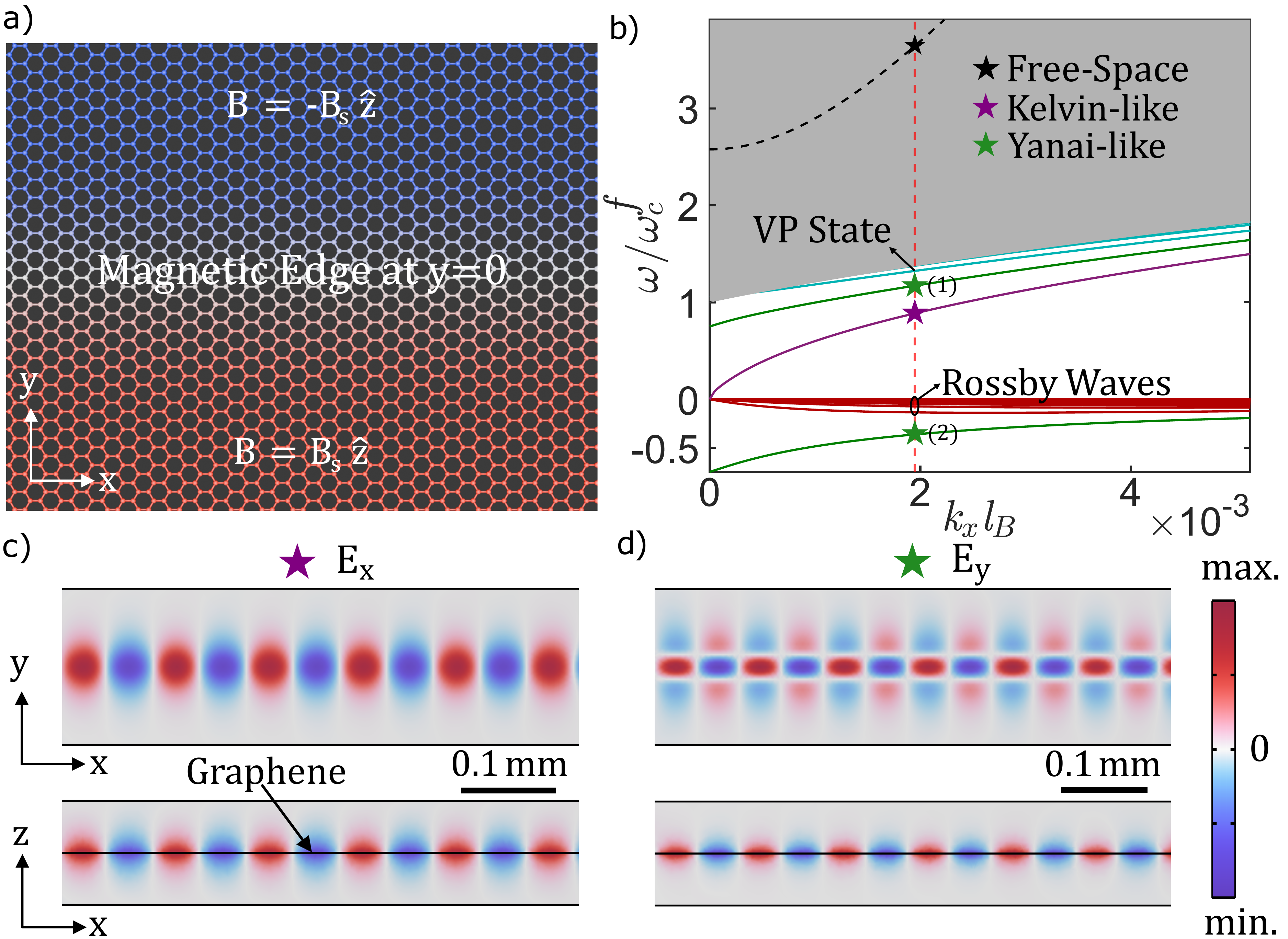}
	\caption{\textbf{Edge Mode Dispersion in Graphene at a Domain Edge.}
		\textbf{a}. Free-standing graphene with a smooth domain edge running along the $y=0$ plane with $\textbf{B}=\textrm{B}_s$ ($-\textrm{B}_s$)$\hat{z}$ in the $y\ll0$ ($y\gg0$) region. \textbf{b}. Edge magneto-plasmon dispersion in graphene; the green (purple) star indicates a topologically protected Yanai-like (Kelvin-like) edge state, where the superscript (1)/(2) represents the Yanai-like wave with a positive/negative phase velocity, and the black star indicates the free-space wave; the black dashed curve shows the air light line with a $k_z\neq0$ offset. \textbf{c}. $\mathrm{E}_x$ field of the Kelvin-like edge state (purple star) localized on the domain edge at $y=0$ in the $x$-$y$ and $x$-$z$ planes, plotted using eigenmode simulations. \textbf{d}. $\mathrm{E}_y$ field of the Yanai-like edge state (green star) localized on the domain edge at $y=0$ in the $x$-$y$ and $x$-$z$ planes, plotted using eigenmode simulations.}
	\label{fig:MT_5} 
\end{figure}

\subsection{Static Electromagnetic Mode in Magnetized Graphene}
\label{Sec:4.2}
Magnetized graphene supports bulk GMPs with a band gap, as shown in the dispersion diagram in Fig.~\ref{fig:MT_4}a. The non-zero-frequency bands (gray) in Fig.~\ref{fig:MT_4}a are topologically nontrivial magneto-plasmon bands, characterized by Chern numbers $C_\pm = \pm 1$ \cite{Mukherjee2025}. In contrast, the zero-frequency mode is static, spatially varying, and topologically trivial. However, unlike in unmagnetized graphene, the static mode in magnetized graphene ($\textrm{B}_s\neq0$) comprises both static electric and magnetic field components. The background magnetic field results in a tensorial magneto-optical conductivity for graphene, hybridizing the magneto-plasmon modes. Thus, the modes cannot be classified as purely transverse magnetic (TM) or transverse electric (TE). Consequently, any polarization of the incident FS wave excites, at a temporal interface, all possible plasmon modes at a fixed $k_x$ (with $k_y=0$). We choose the same mixed-polarization incident wave as in the unmagnetized graphene case. The top (middle) panel of Fig.~\ref{fig:MT_4}b shows the static $\mathrm{B}_x$ ($\mathrm{E}_z$) field of the static electromagnetic mode (pink star) and the propagating bulk GMPs (green and purple stars) at 20 ps, excited at a TI at 6.5 ps; the inset shows the circulating magnetic field lines (black cones) in an enlarged view of the frozen wave. Freely streaming electrons support the static electromagnetic mode in magnetized graphene by creating a spatially varying, static current density $J_y(x)$, as shown in the bottom panel of Fig.~\ref{fig:MT_4}b. The spatially varying electric and magnetic field components in the static electromagnetic mode interact with the external bias magnetic field so that the net Lorentz force on the streaming electrons is zero. In Fig.~\ref{fig:MT_4}c, the red-shifted peak in the FT of the signal at the probe point at $\omega/\omega_c^f\approx1.4$ corresponds to the temporally scattered bulk GMPs excited at a TI with a fixed wave vector parallel to graphene, as shown by the purple and green stars in Fig.~\ref{fig:MT_4}a and b.

\subsection{Free-Space Coupling to Topological Edge States at a Domain Edge}
\label{Sec:4.3}
We consider a magnetic domain edge where the external bias magnetic field ($\mathrm{B_s(y)}$) flips sign. Fig.~\ref{fig:MT_5}a shows a domain edge at $ y=0$, with bulk GMP bands having opposite Chern numbers in the two domains on either side of the edge. The difference in Chern numbers of the bulk magneto-plasmon bands as we move from $y\ll0$ to $y\gg0$ is $C(-\textrm{B}_s)-C(\textrm{B}_s)=2$. The bulk-edge correspondence predicts that two topologically protected edge states, Kelvin-like and Yanai-like edge GMPs that span the bulk GMP gap and are unidirectional, may exist at this domain edge \cite{Tauber2019EqWav, Mukherjee2025}. In addition to topological edge states, a smooth domain edge also supports VP edge states and Rossby waves, which are non-topological edge states that exist in the bulk GMP band gap but neither span the gap nor are unidirectional \cite{Mukherjee2025}. Fig.~\ref{fig:MT_5}b shows the dispersion for both topological and non-topological edge GMPs localized at the domain edge. The purple (green) curve shows the dispersion of the topological Kelvin-like (Yanai-like) edge state. The cyan and dark red curves show the dispersion of the non-topological VP and Rossby GMPs, respectively. Fig.~\ref{fig:MT_5}c,d show the eigenmode simulations of the $\mathrm{E}_x$ and $\mathrm{E}_y$ field profiles in the $x$-$y$ and $x$-$z$ planes for the Kelvin- and Yanai-like topological edge states localized at $y=0$, respectively.
\begin{figure}[t!]
	\centering
	\includegraphics[width=1\textwidth]{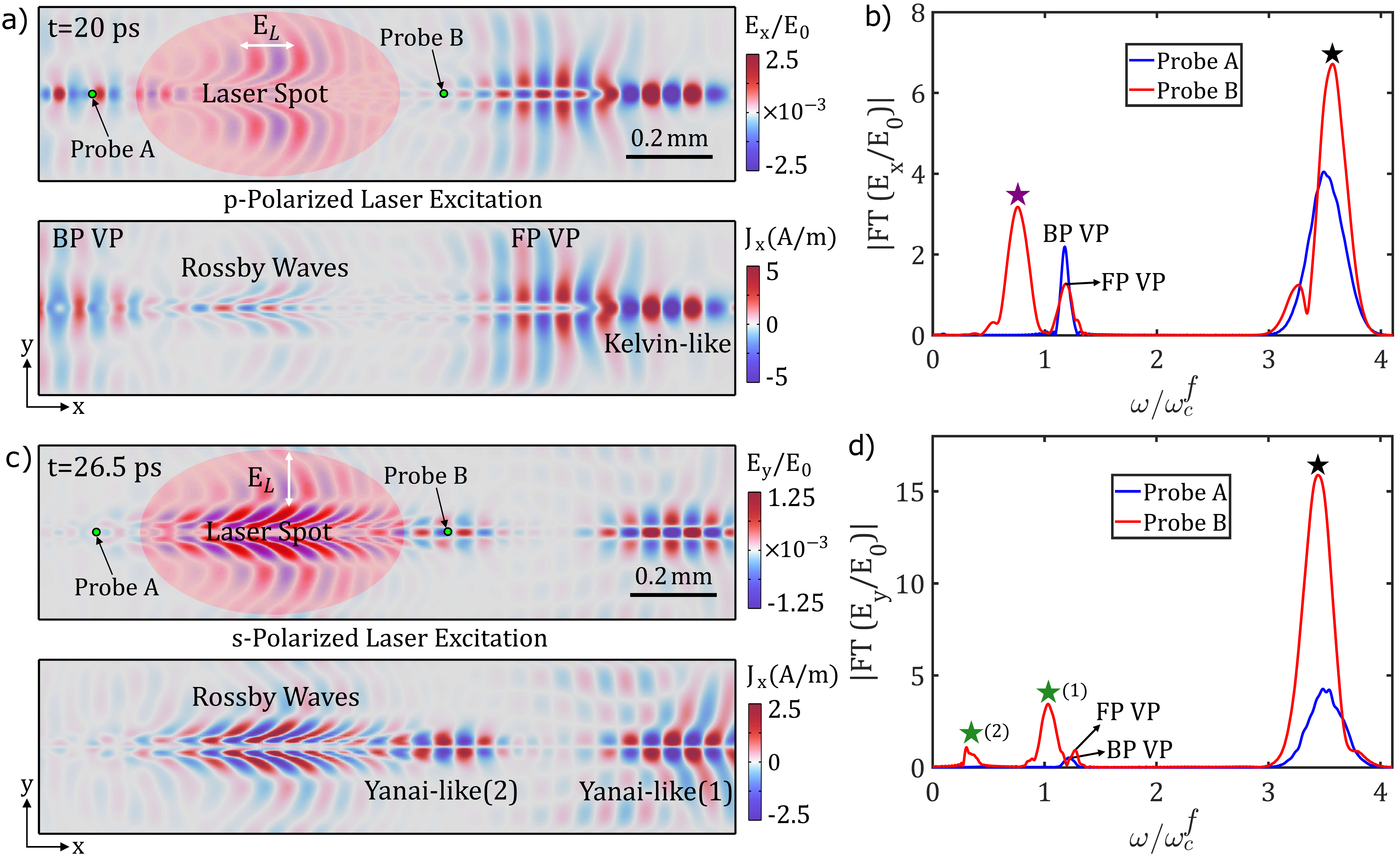}
	\caption{\textbf{Free-Space Excitation of Topological Edge States.}
		\textbf{a}. The top panel shows the $\mathrm{E}_x$ fields of the FP Kelvin-like mode, VP states, and frozen Rossby waves at 20 ps, while the bottom panel shows the x-component of the current density on graphene, excited by a p-polarized Gaussian pulse with a TI at 6.5 ps. \textbf{b}. Spectrum of the $\mathrm{E}_x$ field recorded at the two probe locations shown in $\textbf{a}$, showing the conversion of free-space (FS) light to FP Kelvin-like mode along with VP states. \textbf{c}. The top panel shows the $\mathrm{E}_y$ fields of the FP Yanai-like (1) mode, Yanai-like (2) mode, and frozen Rossby waves at 26.5 ps, while the bottom panel shows the x-component of the current density on graphene, excited by an s-polarized Gaussian pulse with a TI at 6.5 ps. \textbf{d}. Spectrum of the $\mathrm{E}_y$ field recorded at the two probe locations shown in $\textbf{c}$, showing the conversion of free-space (FS) light into FP Yanai-like modes, along with FP and BP VP states.}
	\label{fig:MT_6} 
\end{figure}

For TI simulations, a Gaussian wave with a 1 ps pulse width, centered at 3 ps, was incident at $\theta=45^\circ$ relative to the normal of the graphene plane. We introduced the TI at 6.5 ps, when the pulse reaches the focus, shown as the red shaded region marked \enquote{Laser spot} in the top panels in Figs.~\ref{fig:MT_6}a and c. Of the two topological edge modes, a p-polarized wave excites only the Kelvin-like GMP, along with non-topological VP states, Rossby waves, and bulk GMPs. The top (bottom) panel of Fig.~\ref{fig:MT_6}a shows the $\mathrm{E}_x$-field ($J_x$ current density) at 20 ps for the forward-propagating (FP) Kelvin-like edge mode, forward- and backward-propagating VP states, and localized Rossby waves in the $x$-$y$ plane. Fig.~\ref{fig:MT_6}b shows the $\mathrm{E}_x$-field spectra recorded at the two probe locations in Fig.~\ref{fig:MT_6}a. The peak at $\omega/\omega_c^f \approx 0.8$ for Probe B corresponds to the FP Kelvin-like wave. At the same time, Probe A shows no temporally backscattered signal for the Kelvin-like mode, indicating robust unidirectional propagation of the topologically protected Kelvin-like edge GMP. However, both probes record a signal at $\omega/\omega_c^f\approx1.2$ corresponding to the FP and BP VP edge states, confirming that VP states are non-topological and therefore not immune to backscattering. In contrast, an s-polarized wave excites only the topologically protected FP Yanai-like edge GMP along with non-topological FP and BP VP edge GMPs, Rossby waves, and bulk GMPs at the time interface. The top (bottom) panel of Fig.~\ref{fig:MT_6}c shows the $\mathrm{E}_y$-field ($J_x$ current density) profile of the TI-excited modes at 26.5 ps in the $x$-$y$ plane corresponding to the Yanai-like modes and Rossby waves, where (1)/(2) represent the positive/negative phase-velocity Yanai-like modes. Fig.~\ref{fig:MT_6}d shows the $\mathrm{E}_y$-field spectra recorded at the two probe locations. The peaks at $\omega/\omega_c^f\approx0.35$ and $\omega/\omega_c^f\approx1.1$ correspond to the forward-propagating Yanai-like wave with negative and positive phase velocities, respectively. The absence of a backward-propagating edge state corresponding to the Yanai-like mode at Probe A indicates its immunity to backscattering. Forward- and backward-propagating plasmon signals corresponding to the FP and BP VP edge states are also observed. 

\section{Discussion}
\label{Sec:5}
Time-varying graphene offers an optically controlled active platform for investigating novel time-varying phenomena in two-dimensional materials. Rapidly pumping valence-band electrons into the conduction band creates a transient population of high-energy, non-thermal electrons that strongly scatter electromagnetic waves during photoexcitation. Using an electron-transport model, we derive the thermal and non-thermal current densities from kinetic theory for relativistic electrons in graphene under non-equilibrium electron dynamics. 

During photoexcitation in unmagnetized graphene, a mixed-polarization Gaussian wave excites propagating GPs and a static magnetic-field mode supported by freely streaming electrons perpendicular to the plane of incidence. The transient non-thermal phase of high-energy electrons significantly alters the scattered GP amplitude, with higher-energy electrons producing stronger GPs. Higher-energy non-thermal electrons heat the thermal population and lower the chemical potential, reducing the Drude weight and redshifting the TI-excited plasmon. Furthermore, in magnetized graphene, the background magnetic field hybridizes the plasmon modes, resulting in polarization-agnostic excitation of all possible GMP modes at a TI for a fixed wave vector parallel to the graphene plane. For the same mixed-polarization incident wave, the TI excited propagating bulk plasmons and a static electromagnetic mode with non-zero electric and magnetic components, supported by transversely streaming free electrons. We emphasize that the FS excitation of this static, zero-frequency mode is highly compelling, both fundamentally and for potential applications in photonics and electronics, such as metal-gate-free and optoelectronic-device-free local manipulation of graphene electron density to create graphene-based photonic crystals \cite{Basov2019NatComms}, waveguide channels for surface plasmons \cite{Fang2013APL}, and laser-induced two-dimensional photonic memory \cite{Naomi2015NanoLett, Kim2013APL}, among others.

Additionally, we demonstrated a novel FS coupling mechanism to topological edge states using TIs, eliminating the need for near-field excitation techniques and diffractive gratings to excite these highly localized modes. Notably, the unidirectional propagation of topological edge states prohibits their simultaneous excitation and detection using established scanning probe techniques \cite{Dapolito2023}. During ultrafast photoexcitation, a p- or s-polarized Gaussian wave excited a Kelvin- or Yanai-like topological edge state, respectively, as well as forward- and backward-propagating VP states and Rossby waves along the domain edge. We observed no backward signal corresponding to topological edge states, indicating complete immunity to backscattering. However, non-topological edge states, such as VP states, scatter into both the forward- and backward-propagating waves. This robust protection arises from the non-trivial topology of the bulk bands, with bulk-edge correspondence forcing monotonic group-velocity dispersion along a domain edge and thus preventing backscattering. Lastly, these findings may pave the way for further study of non-equilibrium carrier dynamics in time-varying two-dimensional materials and open new research directions in time-varying two-dimensional photonics.

\subsection* {Disclosures}
The authors declare no competing interests.

\subsection* {Code and Data}
The data that support the findings of this study are available from the corresponding authors upon request.

\subsection* {Acknowledgments}
The work at Cornell was supported by the University of Dayton Research Institute (UDRI) under contract FA8651-24-F-B013 and by the Office of Naval Research (ONR) under grant no. N00014-21-1-2056, and the Army Research Office (ARO) under the award W911NF2110180. Authors J.A. and M.A. acknowledge funding for CLAWS from the Office of the Under Secretary of Defense for Research and Engineering, Applied Research for the Advancement of S\&T Priorities (ARAP) Program. All authors acknowledge Prof. Charles Seyler and Dr. Steven Lantz for providing access to computer resources at the Laboratory of Plasma Studies, Cornell University, where three-dimensional COMSOL time-domain simulations were performed.


\bibliography{main}   
\bibliographystyle{ieeetr}

\end{document}


\maketitle

\renewcommand{\thesection}{S\arabic{section}}
\renewcommand{\thefigure}{S\arabic{figure}}
\renewcommand{\thetable}{S\arabic{table}}
\renewcommand{\refname}{SI References}

\section*{Band Structure Calculations}
The guiding equations for bulk plasmon propagation are the continuity equation, $\partial_t\rho+\nabla\cdot\textbf{J}_s=0$, and the current density equation, $\partial_t\textbf{J}_s=W_D\textbf{E}-\omega_c[\textbf{J}_s\times\hat{\textbf{z}}]$, where $W_D=e^2\mu_f/(\pi\hbar^2)$ is the Drude weight and $\omega_c=e\textrm{B}_sv_F^2/\mu_f$ is the cyclotron frequency. In the following subsections, the GP dispersion is investigated for two cases: i) bulk plasmons in the absence/presence of a background magnetic field, and ii) edge states at the magnetic edge.

\subsection*{Bulk Plasmon Bands}

To calculate the GP dispersion for graphene, we solve the continuity and the current-density equations in Fourier space. Substituting $\textbf{E}=-\nabla\phi$ and $\phi(k)=\rho/[k\epsilon_0(\epsilon_1+\epsilon_2)]$, where $\phi$ is the scalar potential, we rewrite the guiding equations in matrix form, $A|u\rangle=\omega|u\rangle$, where $|u\rangle=(\frac{\alpha}{\sqrt{k}}\rho, J_x, J_y)^T$ and $\alpha=\sqrt{W_D/(2\epsilon_0)}$,
\begin{equation}
    A=\begin{pmatrix}
        0 & \frac{\alpha}{\sqrt{k}}k_x & \frac{\alpha}{\sqrt{k}}k_y \\ \frac{\alpha}{\sqrt{k}}k_x & 0 & -i\omega_c \\ \frac{\alpha}{\sqrt{k}}k_y & i\omega_c & 0
    \end{pmatrix},
\end{equation}
where we take $\epsilon_1=\epsilon_2=1$ (air on both sides for free-standing graphene). The above matrix eigenvalue problem can be solved analytically by calculating the eigenvalues of matrix $A$, which yields the bulk GMP dispersion, given by $\omega_{\pm}(\textbf{k})=\pm\sqrt{\omega_c^2+\alpha^2k}$ (non-zero-frequency valence and conduction plasmon bands), and $\omega_0(\textbf{k})=0$ (zero-frequency plasmon mode). Setting $\textrm{B}_s=0$ T ($\textrm{B}_s=1$ T) recovers the bulk plasmon bands for unmagnetized (magnetized) graphene.

\subsection*{Topological Edge States at a Magnetic Edge}
The plane-wave expansion method (PWEM), as described in \cite{Mukherjee2025}, is used to calculate the topological edge states localized at the magnetic edge of doped graphene. A periodic magnetic field strip along the x-direction is considered, with the magnetic field switching sign twice across the graphene in the y-direction. The magnetic field is represented by two smooth hyperbolic tangent functions multiplied together, $\textbf{B}_s(y)=\textrm{B}_s\mathrm{tanh}\left(\frac{y+w}{l}\right)\mathrm{tanh}\left(-\frac{y-w}{l}\right)\hat{\textbf{z}}$, where $\textrm{B}_s=1$ T, $l=20\ \mu m$, $w=P/4$, and $P=400\ \mu m$ is the periodicity in the y-direction. We rewrite $\rho(y)$, $J_x(y)$, $J_y(y)$, and $\omega_c(y)$ in the plane-wave basis as $A(y)=\sum_{i=-N}^{N}A_i(k_y)e^{ik_{y,i}y}$, where $k_{y,i}=2\pi i/P$, using $2N+1=401$ plane waves in the expansion. Substituting the expanded form into the guiding equations and solving for $\phi(k)$ in the Fourier domain yields the edge-state dispersion. Furthermore, we plot only the edge GMP modes confined to the rising edge of the magnetic periodic strip.

\section*{Boltzmann Transport Equation}
Upon photoexcitation, a symmetric population of non-thermal electrons forms and coexists with a pre-existing symmetric thermal electron population.  However, in the presence of external electromagnetic fields, the probability distribution functions (PDFs) of both electron populations become asymmetric as electrons move collectively in a preferred direction driven by these fields. Nevertheless, the total PDF for graphene electrons, $f_s(\textbf{r},p,\theta,t)$, can be written as the sum of individual PDFs for thermal, $f_T(\textbf{r},p,\theta,t)$, and non-thermal, $f_{NT}(\textbf{r},p,\theta,t)$ electrons, $f_s=f_T+f_{NT}$, where $\textbf{r}\equiv(x,y)$ is the position of the electron on the graphene sheet, $p=|\textbf{p}|$ is the magnitude of the electron momentum, and $\theta$ is the azimuthal angle measured from the x-axis in momentum space. Moreover, assuming isotropic graphene, the analysis is performed far from any graphene spatial edges. Under the assumption of an isotropic and momentum-symmetric photoexcitation source, we expand the total PDF in polar coordinates, writing it as the sum of a symmetric term and asymmetric terms in momentum space: $f_{\alpha}(\textbf{r},p,\theta,t)=f_{\alpha}^{(0)}(p,t)+\delta f_\alpha(\textbf{r}, p,\theta,t)$, where $\delta f_\alpha(\textbf{r},p,\theta,t)=f_{\alpha}^{x}(\textbf{r},p,t)\cos{\theta}+f_{\alpha}^{y}(\textbf{r},p,t)\sin{\theta}$ denotes the first-order perturbation term and $\alpha=T, NT$ represents the electron population type. Here, $f_\alpha^{(0)}$ is symmetric in momentum space, while $\delta f_{\alpha}$ is not. The Boltzmann Transport Equation (BTE) tracks the electron PDF in the presence of electromagnetic fields while accounting for various scattering mechanisms. Since thermal and non-thermal electrons are treated as distinct populations, the BTE is written separately for each. For thermal electrons, the time derivative of $f_{T}(\mathbf{r},p,\theta,t)$ is given by,
\begin{equation}
    \frac{\partial f_{T}}{\partial t}=-\textbf{v}\cdot\nabla_rf_{T}-\textbf{F}_L\cdot\nabla_pf_{T}+\left(\frac{\partial f_{T}}{\partial t}\right)_{e-e}+\left(\frac{\partial f_{T}}{\partial t}\right)_{e-ap}
    \label{eq:df_T_dt1}
\end{equation}
Here, $\left(\frac{\partial f_{T}}{\partial t}\right)_{e-e}$ represents non-thermal-thermal electron elastic collisions, and $\left(\frac{\partial f_{T}}{\partial t}\right)_{e-ap}$ denotes thermal electrons coupling to the background lattice via emission of an acoustic phonon. In the presence of an external electromagnetic field, i.e., $\delta f_{T}(\mathbf{r},p,\theta,t)\propto e^{i\textbf{k}_0\cdot\textbf{r}}$, the first term on the right-hand side of eq.~\ref{eq:df_T_dt1} representing the spatial derivative of the thermal PDF gives: $-i(\mathbf{v}\cdot\mathbf{k}_0)\delta f_{T}=-i(\hat{\mathbf{v}}\cdot\hat{\mathbf{k}}_0)v_Fk_0 \delta f_{T}=-i(\hat{\mathbf{v}}\cdot\hat{\mathbf{k}}_0)\left(\frac{\mu_i}{\hbar}\right)\left(\frac{k_0}{k_i}\right)\delta f_{T}$. Moreover, the free-space wave vector of the driving electromagnetic field, $k_0\approx 10^5\ \mathrm{m}^{-1}$, is considerably smaller than the electronic wave vector near the initial chemical potential, $k_i\approx 10^8\ \mathrm{m}^{-1}$. Since $k_0$ satisfies the long-wavelength limit for the external fields, i.e., $k_0/k_i \ll 1$, the contribution from spatial dispersion is negligible. Therefore, ignoring the non-local effects, the BTE is solved completely in momentum space as:
\begin{equation}   
        \frac{\partial f_{T}}{\partial t}=-\textbf{F}_L\cdot\nabla_pf_{T}+\left(\frac{\partial f_{T}}{\partial t}\right)_{e-e}+\left(\frac{\partial f_{T}}{\partial t}\right)_{e-ap}.
    \label{eq:df_T_dt}
\end{equation}
This implies that the thermal and non-thermal PDFs can be assumed to be independent of electron position and are collectively written as:
\begin{equation}
    f_{\alpha}(p,\theta,t)=f_{\alpha}^{(0)}(p,t)+\delta f_\alpha(p,\theta,t).
    \label{eq:f_Gen}
\end{equation}
Similarly, for non-thermal electrons, the time derivative of the non-thermal PDF is given by 
\begin{equation}
        \frac{\partial f_{NT}}{\partial t}=-\textbf{F}_L\cdot\nabla_pf_{NT}+S_{PE}(p,t)+\left(\frac{\partial f_{NT}}{\partial t}\right)_{e-e},
    \label{eq:df_NT_dt1}
\end{equation}
where $S_{PE}(p,t)$ represents the isotropic photoexcitation source that symmetrically promotes electrons from the valence to the conduction band and hence has no $\theta$ dependence. Here, the collisional time derivative on the right-hand side of eq.~\ref{eq:df_NT_dt1} represents elastic collisions between non-thermal and thermal electrons. Given the short-lived non-thermal phase of photoexcited electrons, we assume single-step relaxation of non-thermal electrons to the thermal sea. Therefore, under the scalar relaxation time approximation, the momentum relaxation time constant during electron-electron elastic collisions is the same as the energy relaxation time constant, i.e., $\left(\frac{\partial [\delta f_{NT}]}{\partial t}\right)_{e-e}=-\frac{\delta f_{NT}}{\tau_\textrm{e-e}}$ and $\left(\frac{\partial f_{NT}^{(0)}}{\partial t}\right)_{e-e}=-\frac{f_{NT}^{(0)}}{\tau_\textrm{e-e}}$, implying $\left(\frac{\partial f_{NT}}{\partial t}\right)_{e-e}=-\frac{f_{NT}}{\tau_\textrm{e-e}}$. Here, $\tau_\textrm{e-e}\approx50$ fs represents the effective ultrafast electron-electron scattering time constant for non-thermal electron relaxation \cite{Polini2013NatComms}. This simplifies eq.~\ref{eq:df_NT_dt1} to:
\begin{equation}
        \frac{\partial f_{NT}}{\partial t}=-\textbf{F}_L\cdot\nabla_pf_{NT}+S_{PE}(p,t)-\frac{f_{NT}}{\tau_\textrm{e-e}}.
    \label{eq:df_NT_dt}
\end{equation}

\section*{Modified Two-Temperature Model}
Graphene in non-equilibrium consists of three interacting species: non-thermal electrons, thermal electrons, and the background lattice. During photoexcitation and collisional absorption of the weak external laser field, the system's total energy increases. The following subsections demonstrate energy conservation by showing that photoexcitation energy transferred to non-thermal electrons relaxes into the thermal sea during ultrafast thermalization, which then transfers some of this energy to the background lattice via acoustic phonon emission. Furthermore, conservation of electron number density after photoexcitation is demonstrated and used to calculate the temporal evolution of the thermal electron density.

\subsection*{Energy Conservation}
The energy density of the system ($\varepsilon_s$) is equal to the sum of the energy densities of i) non-thermal electrons ($\varepsilon_{NT}$), ii) thermal electrons ($\varepsilon_{T}$), and iii) the background lattice ($\varepsilon_{l}$), i.e., $\varepsilon_{s}=\varepsilon_{NT}+\varepsilon_{T}+\varepsilon_{l}$. Since the system gains energy via photoexcitation ($\varepsilon_{PE}$) and collisional absorption of the external laser field ($\varepsilon_L$) by graphene electrons, the rate of change of the system's energy density can be written as: $\partial_t \varepsilon_{s}=\partial_t\varepsilon_{NT}+\partial_t\varepsilon_{T}+\partial_t\varepsilon_{l}=\partial_t \varepsilon_{PE}+\partial_t \varepsilon_L$. The thermal and non-thermal electron energy densities are written as:
\begin{equation}
    \varepsilon_{T}(t)=g_sg_v\int f_{T}(p,\theta,t)E(p)\frac{d^2\textbf{p}}{(2\pi\hbar)^2},\ \varepsilon_{NT}(t)=g_sg_v\int f_{NT}(p,\theta,t)E(p)\frac{d^2\textbf{p}}{(2\pi\hbar)^2}
    \label{eq:E_T_E_NT_1}
\end{equation}
where $g_s=2$ ($g_v=2$) accounts for the electron spin (valley) degeneracy in graphene and $E$ is the electron energy. Taking a partial time derivative of the non-thermal electron energy density and using eq.~\ref{eq:df_NT_dt} gives
\begin{equation}
    \begin{aligned}
    \frac{\partial \varepsilon_{NT}}{\partial t}=-g_sg_v\int \left[E(p)\textbf{F}_L\right]\cdot \nabla_pf_{NT}\frac{d^2\textbf{p}}{(2\pi\hbar)^2}+g_sg_v\int E(p)S_{PE}(p,t)\frac{d^2\textbf{p}}{(2\pi\hbar)^2}\\+g_sg_v\int E(p)\left(-\frac{f_{NT}}{\tau_\textrm{e-e}}\right)\frac{d^2\textbf{p}}{(2\pi\hbar)^2}.
    \label{eq:d_Energy_NT_dt1}
    \end{aligned}
\end{equation}
Simplifying eq.~\ref{eq:d_Energy_NT_dt1} by using the divergence theorem on the first term on the right-hand side yields:
\begin{equation}
    \begin{aligned}
    \frac{\partial \varepsilon_{NT}}{\partial t}=-g_sg_v\oint_{C}f_{NT}E(p)\textbf{F}_L\cdot\hat{\textbf{n}}\frac{dl_p}{(2\pi\hbar)^2} +g_sg_v\int f_{NT}\textbf{F}_L\cdot\nabla_pE(p)\frac{d^2\textbf{p}}{(2\pi\hbar)^2}+\frac{\hbar\omega_p}{2}w_{PE}-\frac{\varepsilon_{NT}}{\tau_\textrm{e-e}},
    \label{eq:d_Energy_NT_dt2}
    \end{aligned}
\end{equation}
where the non-thermal electron energy at the time of photoexcitation is replaced by $E(p)=\hbar\omega_p/2$, and the fact that external fields are independent of electron momentum is used. Here, the line integral over $C$ denotes a closed contour at $p\rightarrow\infty$ with surface normal $\hat{\textbf{n}}$, where $f_{NT}\rightarrow0$ because there are no infinite-energy electrons. Moreover, using the electron dispersion in graphene, $E(p)=v_Fp$, we get, $\nabla_pE=v_F\textbf{p}/p=\textbf{v}$. Further simplification of eq.~\ref{eq:d_Energy_NT_dt2} using the Lorentz force law, $\textbf{F}_L=-e\left[\textbf{E}+\textbf{v}\times \textbf{B}_s\right]$, gives:
\begin{equation}
    \begin{aligned}
    \frac{\partial \varepsilon_{NT}}{\partial t}=\textbf{J}_{NT}\cdot\textbf{E}+\frac{\hbar\omega_p}{2}w_{PE}-\frac{\varepsilon_{NT}}{\tau_\textrm{e-e}}.
    \label{eq:d_Energy_NT_dt3}
    \end{aligned}
\end{equation}
Similarly, the partial time-derivative equation for thermal electron energy density is written as,
\begin{equation}
    \begin{aligned}
    \frac{\partial \varepsilon_{T}}{\partial t}=\textbf{J}_{T}\cdot\textbf{E}+g_sg_v\int E(p)\left(\frac{\partial f_{T}}{\partial t}\right)_{e-e}\frac{d^2\textbf{p}}{(2\pi\hbar)^2}+g_sg_v\int E(p)\left(\frac{\partial f_{T}}{\partial t}\right)_{e-ap}\frac{d^2\textbf{p}}{(2\pi\hbar)^2}.
    \label{eq:d_Energy_T_dt}
    \end{aligned}
\end{equation}
Following energy conservation, i.e., $\partial_t\varepsilon_{NT}+\partial_t\varepsilon_{T}+\partial_t\varepsilon_{l}-\partial_t \varepsilon_{PE}-\partial_t \varepsilon_L=0$, and using eq.~\ref{eq:d_Energy_NT_dt3} and eq.~\ref{eq:d_Energy_T_dt} yields:
\begin{equation}
    \begin{aligned}
    \left[\textbf{J}_{s}\cdot\textbf{E}-\frac{\partial \varepsilon_{L}}{\partial t}\right]+\left[\frac{\hbar\omega_p}{2}w_{PE}-\frac{\partial \varepsilon_{PE}}{\partial t}\right]+\left[g_sg_v\int E(p)\left(\frac{\partial f_{T}}{\partial t}\right)_{e-e}\frac{d^2\textbf{p}}{(2\pi\hbar)^2}-\frac{\varepsilon_{NT}}{\tau_\textrm{e-e}}\right]\\ +\left[g_sg_v\int E(p)\left(\frac{\partial f_{T}}{\partial t}\right)_{e-ap}\frac{d^2\textbf{p}}{(2\pi\hbar)^2}+\frac{\partial \varepsilon_{l}}{\partial t}\right]=0,
    \end{aligned}
    \label{eq:Energy_Cons}
\end{equation}
where $\textbf{J}_s=\textbf{J}_{NT}+\textbf{J}_T$ denotes the total electron current density.

\subsubsection*{Collisional Absorption and Photoexcitation Energy Density}
In eq.~\ref{eq:Energy_Cons}, equating both the first and second bracketed terms on the left-hand side to zero gives the rate of collisional absorption of the laser field and the absorbed photoexcitation energy density as:
\begin{equation}
    \frac{\partial\varepsilon_{L}}{\partial t}=\textbf{J}_s\cdot\textbf{E}\ \mathrm{and}\ \frac{\partial\varepsilon_{PE}}{\partial t}=\frac{\hbar\omega_p}{2}w_{PE},\ \mathrm{respectively.}
\end{equation}

\subsubsection*{Energy Transfer to Lattice}
Furthermore, isolating and setting the fourth bracketed term in eq.~\ref{eq:Energy_Cons} to zero yields:
\begin{equation}
    \frac{\partial \varepsilon_{l}}{\partial t}=-g_sg_v\int E(p)\left(\frac{\partial f_{T}}{\partial t}\right)_{e-ap}\frac{d^2\textbf{p}}{(2\pi\hbar)^2}
\end{equation}
The integral on the right-hand side represents acoustic phonon emission, which maintains thermal equilibrium by enabling heat transfer between the lattice at temperature $T_l$ and thermal electrons at temperature $T_e$. Heat flows from electrons to the lattice when $T_e-T_l>0$ and from the lattice to electrons when $T_e-T_l<0$. According to the two-temperature model in graphene \cite{MacD2009PRL, Viljas2010PRB, ETM2025IOP, GrapheneTTM}, this relationship is expressed as:
\begin{equation}
    \frac{\partial \varepsilon_{l}}{\partial t}=C_l\frac{\partial T_l}{\partial t}=g_{ep}(T_e-T_l),
\end{equation}
Here, $\varepsilon_l=C_lT_l$, where $C_l=5.334\times10^{-4}\ \mathrm{J/(m^2K)}$ is the areal heat capacity of graphene \cite{Xing2017Nanoscale}. Following ultrafast photoexcitation, the electron temperature rises almost instantly; therefore, under the high-temperature approximation, $g_{ep}$ denotes the coupling coefficient between thermal electrons and acoustic phonons, as given by eq.~17 in reference \cite{Viljas2010PRB}: 
\begin{equation}
    g_{ep}=\frac{D^2k_B}{30\pi\rho_m\hbar^5v_F^6}\left[15\mu^4+30\pi^2\mu^2(k_BT_e)^2+7\pi^4(k_BT_e)^4\right],
\end{equation}
where $\rho_m=7.6\times10^{-7}\ \mathrm{kg/m^2}$ is the surface mass density of graphene, $D=19$ eV is the screened deformation potential, and $\mu$ is the chemical potential of thermal electrons.

\subsubsection*{Electron Energy Density during Elastic Collisions}
Following the energy density rate equations discussed above, the simplified eq.~\ref{eq:Energy_Cons} gives the condition on the rate of change of thermal and non-thermal electron energy densities during elastic collisions as:
\begin{equation}
    g_sg_v\int E(p)\left(\frac{\partial f_{T}}{\partial t}\right)_{e-e}\frac{d^2\textbf{p}}{(2\pi\hbar)^2}-\frac{\varepsilon_{NT}}{\tau_\textrm{e-e}}=0,
    \label{eq:NT_T_Energy}
\end{equation}
which is also the mathematical form of the conservation of electron energy during elastic collisions between the non-thermal and thermal populations. Using eq.~\ref{eq:f_Gen} simplifies eq.~\ref{eq:NT_T_Energy} to:
\begin{equation}
    g_sg_v\int E(p)\left(\frac{\partial f_{T}^{(0)}}{\partial t}\right)_{e-e}\frac{d^2\textbf{p}}{(2\pi\hbar)^2}=\frac{\varepsilon_{NT}}{\tau_\textrm{e-e}}.
    \label{eq:f_T_Iso}
\end{equation}
Eq.~\ref{eq:f_T_Iso} contains no first-order corrections because the sine/cosine terms vanish when integrated over momentum space, i.e., $\int_0^{2\pi}\sin(\theta)d\theta=\int_0^{2\pi}\cos(\theta)d\theta=0$. Furthermore, we assume a weak driving laser field and ignore heating due to the $\textbf{J}_{NT}\cdot\textbf{E}$ and $\textbf{J}_{T}\cdot\textbf{E}$ terms in eq.~\ref{eq:d_Energy_NT_dt3} and eq.~\ref{eq:d_Energy_T_dt}, respectively, to obtain the temporal evolution of the non-thermal and thermal electron energy densities:
\begin{equation}
    \frac{\partial \varepsilon_{NT}}{\partial t}=\frac{\hbar\omega_{p}}{2}w_{PE}-\frac{\varepsilon_{NT}}{\tau_\textrm{e-e}}\ \mathrm{and}\ \frac{\partial \varepsilon_{T}}{\partial t}=-g_{ep}(T_e-T_l)+\frac{\varepsilon_{NT}}{\tau_\textrm{e-e}},
    \label{eq:d_E_NT_T_dt}
\end{equation}
where the non-thermal and thermal electron energy densities, using eq.~\ref{eq:E_T_E_NT_1} and eq.~\ref{eq:f_Gen}, are written as
\begin{equation}
    \varepsilon_{NT}(t)=\int_{\mu(t)}^{\infty} f_{NT}^{(0)}(E,t)g(E)EdE\ \mathrm{and}\ \varepsilon_{T}(t)=\int_{0}^{\infty} f_{T}^{(0)}(E,t)g(E)EdE
    \label{eq:E_T_E_NT_2}
\end{equation}
in the electron energy space. Here, $g(E)=\frac{2E}{\pi\hbar^2v_F^2}$ is the electronic density of states in graphene's conduction band, and $E$ is the electron energy measured from the conduction band minimum at the Dirac point. In eq.~\ref{eq:E_T_E_NT_2}, the lower integration limit for non-thermal electrons is deliberately set to $\mu(t)$ in accordance with the Pauli exclusion principle. Thermal electrons exist at the same temperature, $T_e(t)$; hence, the symmetric term in $f_{T}$ follows the Fermi-Dirac distribution function, $f_T^{(0)}(E,t)=\left[1+\exp{\left(\frac{E-\mu(t)}{k_BT_e(t)}\right)}\right]^{-1}$.

\subsection*{Electron Number Conservation}
\begin{figure}
	\centering
	\includegraphics[width=1\textwidth]{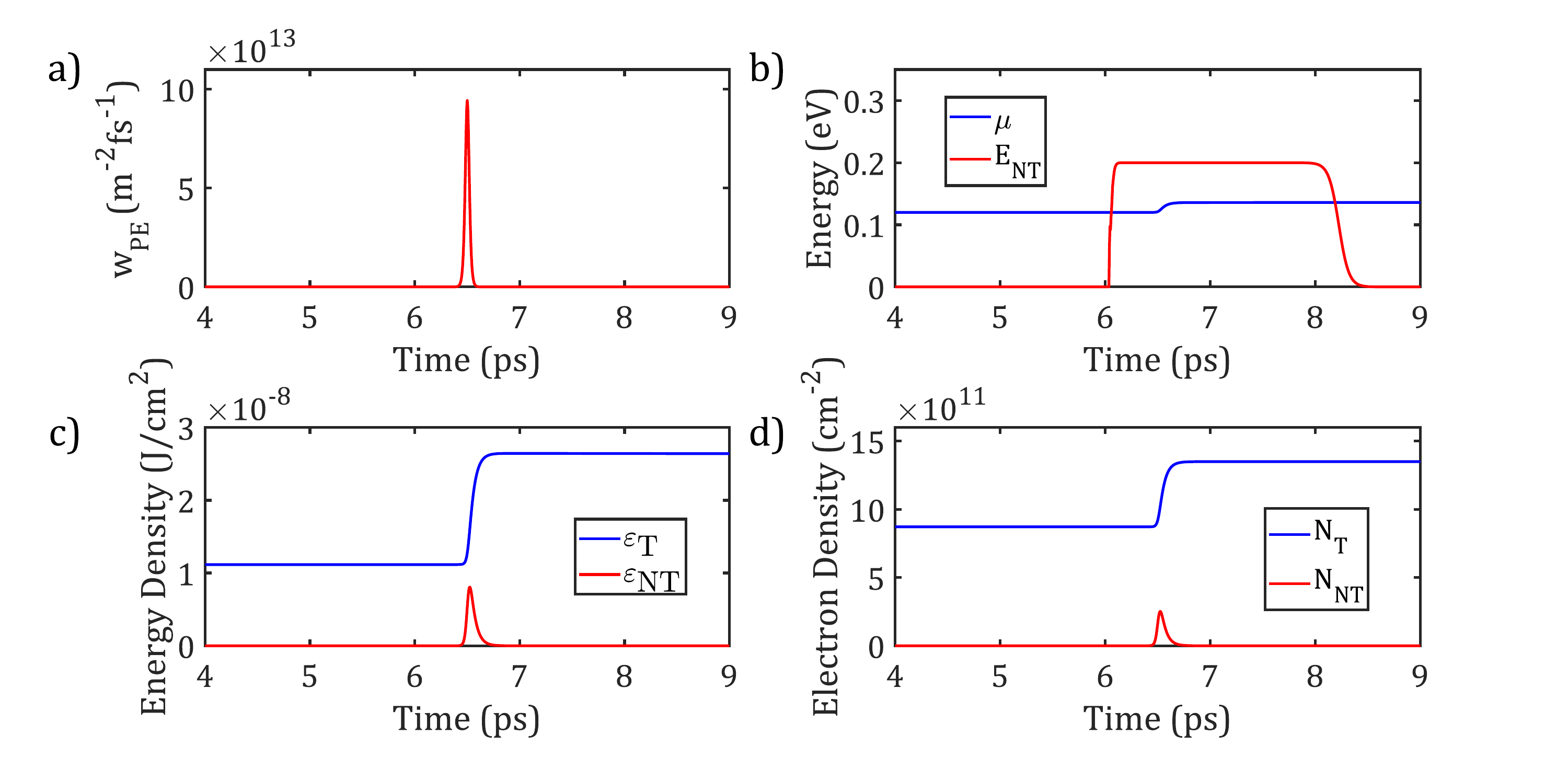}
	\caption{\textbf{Modified TTM for Graphene Electrons.}
		\textbf{a}. Photoexcitation rate ($w_{PE}$) across a TI at 6.5 ps, where $\hbar\omega_p=0.4$ eV. \textbf{b}. Chemical potential ($\mu$) and average non-thermal electron energy ($E_{NT}$) as a function of time. \textbf{c}. Thermal (non-thermal) electron energy density, $\varepsilon_{T}$ ($\varepsilon_{NT}$) as a function of time. \textbf{d}. Thermal (non-thermal) electron density, $N_{T}$ ($N_{NT}$) as a function of time.}
	\label{fig:SI} 
\end{figure}

The non-thermal and thermal electron densities are given by:
\begin{equation}
    N_{NT}(t)=g_sg_v\int f_{NT}(p,\theta,t)\frac{d^2\textbf{p}}{(2\pi\hbar)^2}\ \mathrm{and}\ N_{T}(t)=g_sg_v\int f_{T}(p,\theta,t)\frac{d^2\textbf{p}}{(2\pi\hbar)^2}.
\end{equation}
Taking a partial time derivative of the non-thermal electron density and using eq.~\ref{eq:df_NT_dt} and eq.~\ref{eq:f_Gen} gives:
\begin{equation}
    \begin{aligned}
    \frac{\partial N_{NT}}{\partial t}=-g_sg_v\int \textbf{F}_L\cdot \nabla_pf_{NT}\frac{d^2\textbf{p}}{(2\pi\hbar)^2}+g_sg_v\int S_{PE}(p,t)\frac{d^2\textbf{p}}{(2\pi\hbar)^2}+g_sg_v\int \left(-\frac{f_{NT}}{\tau_\textrm{e-e}}\right)\frac{d^2\textbf{p}}{(2\pi\hbar)^2}.
    \end{aligned}
\end{equation}
Applying the divergence theorem to the first integral and simplifying yields:
\begin{equation}
    \begin{aligned}
    \frac{\partial N_{NT}}{\partial t}=w_{PE}-\frac{N_{NT}}{\tau_\textrm{e-e}}.
    \label{eq:dN_NT_dt}
    \end{aligned}
\end{equation}
Similarly, the partial time derivative of the thermal electron density is written as:
\begin{equation}
    \frac{\partial N_{T}}{\partial t}=g_sg_v\int \left(\frac{\partial f_{T}}{\partial t}\right)_{e-e}\frac{d^2\textbf{p}}{(2\pi\hbar)^2}+g_sg_v\int \left(\frac{\partial f_{T}}{\partial t}\right)_{e-ap}\frac{d^2\textbf{p}}{(2\pi\hbar)^2}.
\end{equation}
Following ref.~\cite{MacD2009PRL}, since the acoustic phonon coupling does not alter the thermal electron density, $\int \left(\frac{\partial f_{T}}{\partial t}\right)_{e-ap}\frac{d^2\textbf{p}}{(2\pi\hbar)^2}=0$; applying eq.~\ref{eq:f_Gen} then gives
\begin{equation}
    \frac{\partial N_{T}}{\partial t}=g_sg_v\int \left(\frac{\partial f_{T}^{(0)}}{\partial t}\right)_{e-e}\frac{d^2\textbf{p}}{(2\pi\hbar)^2}.
\end{equation}
The total electron density in the conduction band, $N_s(t)$, remains constant after non-thermal electron generation; hence, we differentiate $N_s(t)=N_T(t)+N_{NT}(t)$ to obtain the partial time derivative of the thermal electron density as $\frac{\partial N_T}{\partial t}=w_{PE}-\frac{\partial N_{NT}}{\partial t}$, where we have used $\frac{\partial N_s}{\partial t}=w_{PE}$ by assuming that the weak external laser field does not create any new carriers in the system. Therefore, the partial time derivative of the thermal electron density is written as
\begin{equation}
    \frac{\partial N_T}{\partial t}=g_sg_v\int \left(\frac{\partial f_{T}^{(0)}}{\partial t}\right)_{e-e}\frac{d^2\textbf{p}}{(2\pi\hbar)^2}=\frac{N_{NT}}{\tau_\textrm{e-e}},
    \label{eq:dN_T_dt}
\end{equation}
where the non-thermal and thermal electron densities are given by
\begin{equation}
    N_{NT}(t)=\int_{\mu(t)}^{\infty} f_{NT}^{(0)}(E,t)g(E)dE\ \mathrm{and}\ N_{T}(t)=\int_{0}^{\infty} f_{T}^{(0)}(E,t)g(E)dE.
    \label{eq:N_T_N_NT}
\end{equation}
Moreover, eq.~\ref{eq:f_T_Iso} and eq.~\ref{eq:dN_T_dt} ensure energy and particle conservation by the symmetric term of the thermal electron collision integral during elastic collisions between non-thermal and thermal electrons as:
\begin{equation}
    \int E(p) \left(\frac{\partial f_{T}^{(0)}}{\partial t}\right)_{e-e}\frac{d^2\textbf{p}}{(2\pi\hbar)^2}=\left(\frac{\varepsilon_{NT}}{N_{NT}}\right)\int \left(\frac{\partial f_{T}^{(0)}}{\partial t}\right)_{e-e}\frac{d^2\textbf{p}}{(2\pi\hbar)^2}.
    \label{eq:T_Iso_Rel}
\end{equation}
Here, eq.~\ref{eq:T_Iso_Rel} suggests that during elastic collisions between non-thermal and thermal electrons, the thermal population gains, on average, an energy equal to the average energy of the non-thermal electrons, $E_{NT}=\varepsilon_{NT}/N_{NT}$. From the functional forms of $\varepsilon_{NT}$ and $N_{NT}$ in eq.~\ref{eq:d_E_NT_T_dt} and eq.~\ref{eq:dN_NT_dt}, respectively, the average non-thermal electron energy is approximately $E_{NT}\approx\hbar\omega_p/2=0.2$ eV as shown in Fig.~\ref{fig:SI}b.

\subsection*{TTM for Time-Varying Graphene}
Previous studies on thin films and metals have proposed a two-temperature model in which non-thermal electrons act as a delayed source for the pre-existing thermal electron population \cite{Alam2016, ThinFilms_TTM1, ThinFilms_TTM2}. Building on earlier works, we derive a modified two-temperature model (TTM) to track the electron energy and number densities of both electronic populations in graphene over time. Following the derivations presented in previous sections, the time derivatives of the energy (number) density in thermal electrons, $\varepsilon_{T}$ ($N_{T}$), and non-thermal electrons, $\varepsilon_{NT}$ ($N_{NT}$), and the time dependence of the lattice temperature ($T_l$), are dictated by
\begin{equation}
    \begin{gathered}
        \frac{\partial \varepsilon_{T}}{\partial t}=-g_{ep}(T_e-T_l)+\frac{\varepsilon_{NT}}{\tau_\textrm{e-e}},\ \frac{\partial N_{T}}{\partial t}=\frac{N_{NT}}{\tau_\textrm{e-e}},\\ \frac{\partial \varepsilon_{NT}}{\partial t}=-\frac{\varepsilon_{NT}}{\tau_\textrm{e-e}}+\frac{\hbar\omega_{p}}{2}w_{PE},\ \frac{\partial N_{NT}}{\partial t}=-\frac{N_{NT}}{\tau_\textrm{e-e}}+w_{PE},\\ C_l\frac{\partial T_l}{\partial t}=g_{ep}(T_e-T_l),
    \end{gathered}
    \label{eq:Mod_TTM_Energy}
\end{equation}
where $T_e$ is the thermal electron temperature, and $w_{PE}$ is the photoexcitation rate at which an external pump pulse oscillating at frequency $\omega_p$ pumps electrons from the valence to the conduction band. The average non-thermal electron energy is calculated as: $E_{NT} = \varepsilon_{NT}/N_{NT}$, where $\varepsilon_{NT}$ and $N_{NT}$ are calculated using eq.~\ref{eq:Mod_TTM_Energy}. Furthermore, the thermal electron number and energy densities are calculated as: $N_{T}(t)=\int_{0}^{\infty} f_{T}^{(0)}(E,t)g(E)dE\ \mathrm{and}\ \varepsilon_{T}(t)=\int_{0}^{\infty} f_{T}^{(0)}(E,t)g(E)EdE$. The chemical potential ($\mu$) and the thermal electron temperature are determined by equating the thermal electron energy and number densities, as given in eq.~\ref{eq:Mod_TTM_Energy}, to their respective integral forms. Fig.~\ref{fig:SI} shows various TTM parameters calculated using eq.~\ref{eq:Mod_TTM_Energy} for pump photons with energy $\hbar\omega_p=0.4$ eV.

\section*{Time-Varying Graphene Current Density}
The total current density in time-varying graphene equals the sum of the individual current densities of the thermal and non-thermal electron populations. The following subsections derive the time-evolution equations for the individual electron populations using the Boltzmann Transport Equations.

\subsection*{Non-thermal Current Density}
The non-thermal current density is calculated by multiplying the electronic charge by the velocity moment of non-thermal electrons and is written as
\begin{equation}
    \textbf{J}_{NT}(t)=-g_sg_ve\int\textbf{v}f_{NT}(p,\theta,t)\frac{d^2\textbf{p}}{(2\pi\hbar)^2},
    \label{eq:J_NT}
\end{equation}
where $\textbf{v}=\textbf{p}v_F/p$ for graphene electrons. Taking a partial time derivative of eq.~\ref{eq:J_NT} and simplifying using eq.~\ref{eq:df_NT_dt} gives,
\begin{equation}
    \frac{\partial\textbf{J}_{NT}}{\partial t}=g_sg_ve\left[\int\textbf{v}\textbf{F}_L\cdot \nabla_pf_{NT}\frac{d^2\textbf{p}}{(2\pi\hbar)^2}-\int\textbf{v}S_{PE}(p,t)\frac{d^2\textbf{p}}{(2\pi\hbar)^2}+\int\textbf{v}\frac{f_{NT}}{\tau_\textrm{e-e}}\frac{d^2\textbf{p}}{(2\pi\hbar)^2}\right],
    \label{eq:dJ_NT_dt1}
\end{equation}
where the relaxation-time approximation simplifies the third term on the right-hand side of eq.~\ref{eq:dJ_NT_dt1}. The second integral in eq.~\ref{eq:dJ_NT_dt1} vanishes because the source term is assumed isotropic, whereas the last term factors out $\frac{1}{\tau_\textrm{e-e}}$ from the integral. To evaluate the first integral, we simplify the integrand to get $\textbf{v}\textbf{F}_L\cdot \nabla_pf_{NT}=\textbf{v}\left[\nabla_p\cdot\left(f_{NT}\textbf{F}_L\right)-f_{NT}\left(\nabla_p\cdot\textbf{F}_L\right)\right]$. The second term inside the square brackets, upon simplification using $\textbf{v}=v_F\textbf{p}/p=v_F\hat{\textbf{p}}$ and implementing the divergence formula, yields $\left(\nabla_p\cdot\textbf{F}_L\right)=-e\left[\nabla_p\cdot\textbf{E}+v_F\textbf{B}_s\cdot\left(\nabla_p\times\hat{\textbf{p}}\right)-v_F\hat{\textbf{p}}\cdot\left(\nabla_p\times\textbf{B}_s\right)\right]=0$ as the external electromagnetic fields do not depend on the electron's momentum and $\nabla_p\times\hat{\textbf{p}}=0$ for any vector $\textbf{p}$. Therefore, eq.~\ref{eq:dJ_NT_dt1} simplifies to
\begin{equation}
        \frac{\partial\textbf{J}_{NT}}{\partial t}=g_sg_ve\int\textbf{v}\nabla_p\cdot\left(f_{NT}\textbf{F}_L\right)\frac{d^2\textbf{p}}{(2\pi\hbar)^2}+\frac{g_sg_ve}{\tau_\textrm{e-e}}\int\textbf{v}f_{NT}\frac{d^2\textbf{p}}{(2\pi\hbar)^2}.
    \label{eq:dJ_NT_dt2}
\end{equation}
Rewriting the integrand inside the integral term on the right side of eq.~\ref{eq:dJ_NT_dt2} component-wise yields $v_i\nabla_p\cdot\left(f_{NT}\textbf{F}_L\right)=\nabla_p\cdot\left(f_{NT}v_i\textbf{F}_L\right)-f_{NT}\textbf{F}_L\cdot\nabla_pv_i$, where $i=x,y$. Upon further simplification using the divergence theorem, eq.~\ref{eq:dJ_NT_dt2} yields:
\begin{equation}
        \frac{\partial \textrm{J}_{NT,i}}{\partial t}=g_sg_ve\left[\oint_C\left(f_{NT}v_i\textbf{F}_L\cdot\hat{\textbf{n}}\right)\frac{dl_p}{(2\pi\hbar)^2}-\int f_{NT}\textbf{F}_L\cdot\nabla_pv_i\frac{d^2\textbf{p}}{(2\pi\hbar)^2}+\frac{1}{\tau_\textrm{e-e}}\int v_i f_{NT}\frac{d^2\textbf{p}}{(2\pi\hbar)^2}\right],
    \label{eq:dJ_NT_dt3}
\end{equation}
where $\hat{\textbf{n}}$ is the unit vector normal to the closed-curve boundary $C$, and $dl_p$ is the infinitesimal arc-length element along the momentum boundary as $p\rightarrow \infty$. The first integral on the right side of eq.~\ref{eq:dJ_NT_dt3} goes to zero as $p\rightarrow \infty$ because there are no electrons with infinite energy ($f_{NT}|_{p\rightarrow\infty}=0$). Using $\textbf{v}/v_F=\textbf{p}/p$ for graphene electrons, we solve the second integral term as $\nabla_p\textbf{v}=v_F\nabla_p[\textbf{p}/p]=[\textbf{I}-\hat{\textbf{p}}\hat{\textbf{p}}]/m_e(p)$, where $\textbf{I}$ is the identity matrix and $m_e(p)=p/v_F=E/v_F^2$ is the electron effective mass. Further simplifying the integrand in the second integral by using the Lorentz force law yields: $\textbf{F}_L\cdot\nabla_p\textbf{v}=-\left(ev_F^2/E(p)\right)\left[\textbf{E}+\textbf{v}\times\textbf{B}_s-\left(\textbf{v}\textbf{v}\right)\cdot\textbf{E}/v_F^2\right]=\textbf{a}$, where $\textbf{a}$ is the relativistic acceleration of a non-thermal electron moving with momentum $\textbf{p}$. The expanded form of the partial time derivative of $\textbf{J}_{NT}$ in eq.~\ref{eq:dJ_NT_dt3} is written as:
\begin{equation}
    \begin{aligned}
        \frac{\partial\textbf{J}_{NT}}{\partial t}=g_sg_ve^2v_F^2\textbf{E}\int \frac{f_{NT}}{E(p)}\frac{d^2\textbf{p}}{(2\pi\hbar)^2}+g_sg_ve^2v_F^2\left(\int\frac{f_{NT}}{E(p)}\textbf{v}\frac{d^2\textbf{p}}{(2\pi\hbar)^2}\right)\times\textbf{B}_s\\-g_sg_ve^2\left(\int\frac{f_{NT}}{E(p)}\textbf{vv}\frac{d^2\textbf{p}}{(2\pi\hbar)^2}\right)\cdot\textbf{E}+\frac{g_sg_ve}{\tau_\textrm{e-e}}\int f_{NT}\textbf{v}\frac{d^2\textbf{p}}{(2\pi\hbar)^2}.
    \end{aligned}
    \label{eq:dJ_NT_dt4}
\end{equation}
Since graphene has sufficient density of states at $\hbar\omega_p/2$ to support all photoexcited carriers, non-thermal electrons exist only in a highly localized, narrow energy band that relaxes in a single step to the thermal sea (see Fig.~\ref{fig:SI}b). Therefore, $E(p)\approx E_{NT}=\hbar\omega_p/2$ can be treated as a constant and factored out of the integrals on the right-hand side of eq.~\ref{eq:dJ_NT_dt4}. Furthermore, using eq.~\ref{eq:f_Gen}, the integrals in eq.~\ref{eq:dJ_NT_dt4} are simplified as
\begin{equation}
    \begin{aligned}
        \frac{g_sg_v}{E_{NT}}\int f_{NT}^{(0)}\frac{d^2\textbf{p}}{(2\pi\hbar)^2}=\frac{N_{NT}}{E_{NT}},\ \ \ \frac{g_sg_v}{E_{NT}}\int \delta f_{NT}\textbf{v}\frac{d^2\textbf{p}}{(2\pi\hbar)^2}=-\frac{\textbf{J}_{NT}}{eE_{NT}},\\ \frac{g_sg_v}{E_{NT}}\int f_{NT}^{(0)}\textbf{vv}\frac{d^2\textbf{p}}{(2\pi\hbar)^2}=\frac{N_{NT}v_F^2}{2E_{NT}}\textbf{I},\ \mathrm{and} \ g_sg_v\int \delta f_{NT}\textbf{v}\frac{d^2\textbf{p}}{(2\pi\hbar)^2}=-\frac{\textbf{J}_{NT}}{e},
    \end{aligned}
    \label{eq:NT_Integrals}
\end{equation}
where $\textbf{v}\textbf{v}=v_F^2\begin{pmatrix}\cos^2{\theta} & \cos{\theta}\sin{\theta}\\ \sin{\theta}\cos{\theta} & \sin^2{\theta}\end{pmatrix}$ is expanded in polar coordinates to evaluate the first integral in the second row of eq.~\ref{eq:NT_Integrals} and $\textbf{I}$ is the identity matrix. Replacing these four integrals in eq.~\ref{eq:dJ_NT_dt4}, we get
\begin{equation}
        \frac{\partial\textbf{J}_{NT}}{\partial t}=\frac{e^2v_F^2N_{NT}}{2E_{NT}}\textbf{E}-\frac{ev_F^2}{E_{NT}}\textbf{J}_{NT}\times\textbf{B}_s-\frac{\textbf{J}_{NT}}{\tau_\textrm{e-e}}.
    \label{eq:dJ_NT_dt5}
\end{equation}

\subsection*{Non-thermal Momentum Relaxation}
Non-thermal electrons lose momentum through elastic collisions with thermal electrons. Furthermore, electron-electron collisions conserve total electronic momentum by redistributing it within the electronic system. Therefore, the non-thermal momentum transferred to the thermal population using eq.~\ref{eq:f_Gen} is written as,
\begin{equation}
    \left(\Delta P_{NT}\right)_{ee}=g_sg_v\int\textbf{p}\left(\frac{\partial f_{NT}}{\partial t}\right)_{e-e}\frac{d^2\textbf{p}}{(2\pi\hbar)^2}=g_sg_v\int\left(\frac{E(p)}{v_F^2}\right)\textbf{v}\left(\frac{\partial [\delta f_{NT}]}{\partial t}\right)_{e-e}\frac{d^2\textbf{p}}{(2\pi\hbar)^2}.
    \label{eq:P_lost}
\end{equation}
On the other hand, the gain in thermal momentum using eq.~\ref{eq:f_Gen} is written as,
\begin{equation}
    \left(\Delta P_{T}\right)_{ee}=g_sg_v\int\textbf{p}\left(\frac{\partial f_{T}}{\partial t}\right)_{e-e}\frac{d^2\textbf{p}}{(2\pi\hbar)^2}=g_sg_v\int\left(\frac{E(p)}{v_F^2}\right)\textbf{v}\left(\frac{\partial [\delta f_{T}]}{\partial t}\right)_{e-e}\frac{d^2\textbf{p}}{(2\pi\hbar)^2}.
    \label{eq:P_Gain}
\end{equation} 
By momentum conservation, the gain in thermal momentum must equal the non-thermal momentum lost via electron-electron collisions. Therefore, equating the change in electronic momenta during electron-electron elastic collisions to zero gives:
\begin{equation}    
    g_sg_v\int E(p)\textbf{v}\left(\frac{\partial [\delta f_{T}]}{\partial t}\right)_{e-e}\frac{d^2\textbf{p}}{(2\pi\hbar)^2}+g_sg_v\int E(p)\textbf{v}\left(\frac{\partial [\delta f_{NT}]}{\partial t}\right)_{e-e}\frac{d^2\textbf{p}}{(2\pi\hbar)^2}=0.
\label{eq:p_connection1}
\end{equation}
Simplifying the non-thermal integral using eq.~\ref{eq:NT_Integrals} and the relaxation time approximation gives 
\begin{equation}    
    g_sg_v\int E(p)\textbf{v}\left(\frac{\partial [\delta f_{T}]}{\partial t}\right)_{e-e}\frac{d^2\textbf{p}}{(2\pi\hbar)^2}=-E_{NT}\left(\frac{\textbf{J}_{NT}}{e\tau_\textrm{e-e}}\right).
\label{eq:p_connection2}
\end{equation}
Furthermore, we simplify eq.~\ref{eq:p_connection2} by assuming that only the electrons near the chemical potential contribute significantly to the thermal electron conductivity, i.e., $E(p)=\mu$ on the left-hand side, and use eq.~\ref{eq:f_Gen} to get 
\begin{equation}    
    g_sg_v\int\textbf{v}\left(\frac{\partial [\delta f_{T}]}{\partial t}\right)_{e-e}\frac{d^2\textbf{p}}{(2\pi\hbar)^2}=-\left(\frac{E_{NT}}{\mu}\right)\left(\frac{\textbf{J}_{NT}}{e\tau_\textrm{e-e}}\right).
\label{eq:f_T_Aniso}
\end{equation}
Eq.~\ref{eq:f_T_Aniso} specifies the electron-momentum dependence of the asymmetric term in the thermal electron collision integral arising from momentum exchange between non-thermal and thermal electrons during elastic collisions.

\subsection*{Thermal Current Density}
The current density of thermal electrons is calculated by multiplying the electronic charge by the velocity moment of thermal electrons and is written as,
\begin{equation}
    \textbf{J}_{T}(t)=-g_sg_ve\int\textbf{v}f_{T}(p,\theta,t)\frac{d^2\textbf{p}}{(2\pi\hbar)^2},
    \label{eq:J_T}
\end{equation}
Taking the time derivative of the thermal current density and using eq.~\ref{eq:df_T_dt} gives,
\begin{equation}
    \frac{\partial \textbf{J}_{T}}{\partial t}=g_sg_ve\left[\int\textbf{v}\textbf{F}_L\cdot \nabla_{p}f_{T}\frac{d^2\textbf{p}}{(2\pi\hbar)^2}-\int\textbf{v}\left(\frac{\partial f_{T}}{\partial t}\right)_{e-e}\frac{d^2\textbf{p}}{(2\pi\hbar)^2}-\int\textbf{v}\left(\frac{\partial f_{T}}{\partial t}\right)_{e-ap}\frac{d^2\textbf{p}}{(2\pi\hbar)^2}\right].
    \label{eq:dJ_T_dt1}
\end{equation}
Performing a similar algebraic simplification to evaluate the first term on the right-hand side of eq.~\ref{eq:dJ_T_dt1}, as was done for the non-thermal current density and using eq.~\ref{eq:f_Gen}, yields:
\begin{equation}
\begin{aligned}
    \frac{\partial \textbf{J}_{T}}{\partial t}=g_sg_ve^2v_F^2\textbf{E}\int \frac{f_{T}^{(0)}}{E(p)}\frac{d^2\textbf{p}}{(2\pi\hbar)^2}+g_sg_ve^2v_F^2\left(\int\frac{\delta f_{T}}{E(p)}\textbf{v}\frac{d^2\textbf{p}}{(2\pi\hbar)^2}\right)\times\textbf{B}_s\\-g_sg_ve^2\left(\int\frac{f_{T}^{(0)}}{E(p)}\textbf{vv}\frac{d^2\textbf{p}}{(2\pi\hbar)^2}\right)\cdot\textbf{E}-g_sg_ve\int\textbf{v}\left(\frac{\partial \left[\delta f_{T}\right]}{\partial t}\right)_{e-e}\frac{d^2\textbf{p}}{(2\pi\hbar)^2}-\frac{\textbf{J}_T}{\tau_m},
    \label{eq:dJ_T_dt2}
    \end{aligned}
\end{equation}
where the third term on the right-hand side of eq.~\ref{eq:dJ_T_dt1} simplifies to $-\textbf{J}_T/\tau_m$, assuming that thermal electrons undergo momentum relaxation to the lattice with an effective time constant $\tau_m\approx10$ ps, which is significantly shorter than the nanosecond-scale time constant for energy relaxation. This difference arises because the speed of sound in graphene is much lower than the Fermi velocity, implying that each acoustic phonon emission transfers only a few meV of energy but can substantially alter the electron's momentum, up to $\Delta p_\mathrm{max}=2\hbar\sqrt{\pi N_{T}}$. The first three terms on the right-hand side of eq.~\ref{eq:dJ_T_dt2} are solved separately. The first term, in energy space, simplifies to:
\begin{equation}
    g_sg_ve^2v_F^2\textbf{E}\int \frac{f_{T}^{(0)}}{E(p)}\frac{d^2\textbf{p}}{(2\pi\hbar)^2}=e^2v_F^2\textbf{E}\int_0^\infty f_{T}^{(0)}\frac{2}{\pi\hbar^2v_F^2}dE=\frac{2e^2\mu}{\pi\hbar^2}\textbf{E}.
    \label{eq:I_T1}
\end{equation}
Since only the electrons near the chemical potential contribute significantly to the net conduction, the electron energy can be replaced by the chemical potential in the integrals containing terms linear in $\textbf{v}$ in the limit $k_BT_e\ll\mu$. This simplifies the second term on the right-hand side of eq.~\ref{eq:dJ_T_dt2} to:
\begin{equation}
    g_sg_ve^2v_F^2\left(\int\frac{\delta f_{T}}{E(p)}\textbf{v}\frac{d^2\textbf{p}}{(2\pi\hbar)^2}\right)\times\textbf{B}_s=\frac{g_sg_ve^2v_F^2}{\mu}\left(\int\delta f_{T}\textbf{v}\frac{d^2\textbf{p}}{(2\pi\hbar)^2}\right)\times\textbf{B}_s=-\frac{ev_F^2}{\mu}\textbf{J}_{T}\times\textbf{B}_s.
    \label{eq:I_T2}
\end{equation}
Furthermore, the electron velocity is written in polar coordinates, and $\textbf{v}\textbf{v}$ is expanded in matrix form to solve the third term on the right-hand side of eq.~\ref{eq:dJ_T_dt2} as:
\begin{equation}
    -g_sg_ve^2\left(\int\frac{f_{T}^{(0)}}{E(p)}\textbf{vv}\frac{d^2\textbf{p}}{(2\pi\hbar)^2}\right)\cdot\textbf{E}=-\frac{e^2\mu}{\pi\hbar^2}\textbf{I}\cdot\textbf{E}=-\frac{e^2\mu}{\pi\hbar^2}\textbf{E}.
    \label{eq:I_T3}
\end{equation}
Substituting the results from eq.~\ref{eq:I_T1}, eq.~\ref{eq:I_T2}, and eq.~\ref{eq:I_T3} into eq.~\ref{eq:dJ_T_dt2} and replacing the fourth term on the right-hand side of eq.~\ref{eq:dJ_T_dt2} using eq.~\ref{eq:f_T_Aniso} yields:
\begin{equation}
    \frac{\partial \textbf{J}_{T}}{\partial t}=\frac{e^2\mu}{\pi\hbar^2}\textbf{E}-\frac{ev_F^2}{\mu}\textbf{J}_{T}\times\textbf{B}_s+\textbf{J}_{NT}\left(\frac{E_{NT}}{\mu\tau_\textrm{e-e}}\right)-\frac{\textbf{J}_T}{\tau_m}.
    \label{eq:dJ_T_dt3}
\end{equation}

\section*{Numerical Simulations}
Two-dimensional time-domain simulations of frozen waves in bulk graphene were performed using the COMSOL time-domain (ewt) solver on an in-house workstation with 112 threads at 2 GHz and 384 GB of DDR4 system memory at 4800 MT/s. A complete two-dimensional simulation with a simulation time of 30 ps took 5 h 45 min for unmagnetized graphene and 7 h 3 min for magnetized graphene on the workstation using 56 threads. Three-dimensional time-domain simulations of free-space excitation of topological edge states in graphene were performed using COMSOL's ewt solver on the ZEUS workstation of the Laboratory of Plasma Studies at Cornell University. ZEUS runs 128 threads at 2.7 GHz (turbo up to 4 GHz) and has 512 GB of DDR5 RAM running at 4800 MT/s \cite{ZEUSwebsite}. A complete three-dimensional simulation with a simulation time of 30 ps required 35 h 45 min to run on ZEUS using 64 threads.

\bibliographystyle{ieeetr}
\bibliography{supp}